\documentclass[10pt,letterpaper,journal]{IEEEtran}
\usepackage{float}
\usepackage{graphicx}
\usepackage[caption=false,font=footnotesize]{subfig}

\usepackage{amsmath}
\usepackage{booktabs}
\usepackage{textcomp}
\usepackage{url}
\usepackage{amssymb}

\usepackage{cite}
\begin{document}

\title{ReLViC: Loss-Resilient Learned Video Coding with Dispersed Packetization and Controllable Packet Dependencies}

\author{Xuyang Chen,~\IEEEmembership{Student Member,~IEEE,} Daquan Feng,~\IEEEmembership{Member,~IEEE,} Xianfu Chen,~\IEEEmembership{Senior Member,~IEEE,} and Xiang-Gen Xia,~\IEEEmembership{Fellow,~IEEE}
\thanks{Xuyang Chen and Daquan Feng are with the College of Electronics and Information Engineering, Shenzhen University, Shenzhen 518060, China (e-mail: chenxuyang2021@email.szu.edu.cn; fdquan@szu.edu.cn).}
\thanks{Xianfu Chen is with Shenzhen CyberAray Network Technology Company Ltd., Shenzhen 518038, China (e-mail: xianfu.chen@ieee.org).}
\thanks{Xiang-Gen Xia is with the Department of Electrical and Computer Engineering, University of Delaware, Newark, DE 19716 USA (e-mail: xxia@ee.udel.edu).}
}
\maketitle

\begin{abstract}
Packet loss can severely impair learned video coding because missing latent tokens compromise both spatial reconstruction and temporal prediction. We present ReLViC, a loss-resilient learned video coding framework that jointly addresses latent coding and packet-loss recovery. ReLViC disperses spatially adjacent latent tokens across packets and employs a dual-purpose Transformer to estimate entropy-model parameters during coding and reconstruct missing latent tokens at the receiver. It controls packet dependencies through a periodic-reset packet-context topology parameterized by the segment length, thereby tuning the trade-off between compression efficiency and error-propagation range without retraining. A three-stage progressive training procedure establishes single-frame coding, learns temporal context for entropy modeling, and then optimizes the recovery of masked latent tokens under simulated packet loss. Experiments using burst-loss traces evaluate ReLViC against H.265 protected by Reed--Solomon forward error correction (FEC) and GRACE, a loss-resilient learned video codec. ReLViC delivers more stable reconstruction and outperforms both baselines under severe packet loss.
\end{abstract}


\section{Introduction}\label{sec:introduction}

Interactive video services, including video conferencing, cloud gaming, and remote collaboration, require frames to arrive before strict playback deadlines~\cite{AFZAL2023103581,BAENA2023109808}. To meet these deadlines, systems may have to sacrifice visual quality when congestion or wireless channel variations cause packet loss. Burst losses are particularly damaging because they can trigger decoding failures, freezes, and error propagation across subsequent frames~\cite{10.1145/3607139,10668820}.

Conventional video codecs achieve high compression efficiency through tightly coupled prediction, transform coding, quantization, and entropy coding. Advanced Video Coding (AVC), High Efficiency Video Coding (HEVC), and Versatile Video Coding (VVC) exploit spatial and temporal redundancy, with later coding decisions often conditioned on earlier reconstructed syntax elements and reference pictures~\cite{wiegand2003avc,sullivan2012hevc,bross2021vvc}. These dependencies improve efficiency when the bitstream is received intact, but they also make compressed video sensitive to packet loss. A lost packet may damage the current frame and propagate errors to subsequent frames through inter-frame prediction, causing persistent visual artifacts.

Learned video compression replaces many hand-designed coding tools with end-to-end optimized latent representations and probability models. Early end-to-end frameworks jointly optimize motion and residual coding~\cite{lu2019dvc}, after which deep contextual video compression (DCVC) shifts the field toward conditional coding with learned feature-domain contexts~\cite{RN2940}. More recent feature-modulated and efficiency-oriented architectures support wider quality ranges, longer prediction chains, and real-time operation, bringing neural codecs closer to practical video delivery~\cite{10655044,Jia_2025_CVPR}. Their efficiency nevertheless relies on structured latent representations and accurate spatiotemporal context. Entropy decoding and reconstruction generally assume that all required latent tokens are available at the receiver. When a packet is lost, part of the latent representation becomes unavailable, which may degrade reconstruction quality and disrupt the decoding of tokens that depend on the missing information.

Traditional error-control mechanisms have to balance delay against bandwidth. Automatic repeat request (ARQ) and hybrid ARQ (HARQ) react only after loss detection and cannot guarantee recovery before the playback deadline when the remaining delay budget is insufficient~\cite{lin1984arq}. Forward error correction (FEC) sends redundancy proactively, but selecting an effective redundancy level requires predicting the severity and duration of the next loss burst~\cite{RN2195}. Modern interactive streaming systems therefore optimize initial transmissions, retransmissions, and redundancy jointly to reduce deadline misses and bandwidth cost~\cite{meng2024hairpin}. Source partitioning, interleaving, intra refresh, and decoder-side error concealment can limit visual damage, but they either reduce coding efficiency or are tailored to the syntax and tools of particular codecs~\cite{664283,855913,RN2798}.

Recent neural approaches make the codec itself aware of loss. GRACE jointly trains its encoder and decoder over simulated packet losses to obtain graceful degradation~\cite{RN2730}. Reparo performs generative completion from received video tokens to reduce freezes in video conferencing~\cite{RN2680}. NeuralMDC combines correlated multiple descriptions with masked token prediction for unreliable multipath networks~\cite{RN2937}. Morphe uses vision foundation models and a loss-resilient streaming system for high-fidelity generative video delivery~\cite{316606}. At the image level, ResiComp shows that a shared masked Transformer can support both entropy modeling and feature-domain packet-loss recovery~\cite{RN2932}. However, existing approaches do not fully resolve temporal error propagation in video codecs, the trade-off between compression efficiency and loss resilience, rapid adaptation to changing network conditions, or recovery from sustained burst losses.

We develop ReLViC, a loss-resilient learned video coding framework for digital packet networks. Dispersed latent packetization assigns spatially adjacent latent tokens to different packets so that a single packet loss does not create a large contiguous region of missing tokens. A dual-purpose Transformer estimates conditional distributions during coding and reconstructs missing latent tokens using shared context. A periodic-reset context rule controls packet dependencies through the segment length $g$, with a maximum error-propagation span of $g-1$ packets. A three-stage progressive training strategy successively establishes single-frame coding, temporal context modeling, and packet-loss recovery. The main contributions are summarized below.
\begin{itemize}
    \item We propose ReLViC, an end-to-end loss-resilient learned video coding framework that jointly optimizes latent coding, dispersed packetization, entropy modeling, and packet-loss recovery for digital packet networks.
    \item We develop a dual-purpose Transformer and a three-stage progressive training strategy. The shared context model estimates conditional distributions for entropy coding and reconstructs missing latent tokens from received spatial and temporal evidence.
    \item We introduce dispersed latent packetization and controllable packet dependencies. Together, they transform concentrated packet erasures into spatially distributed patterns of missing tokens and allow the codec to balance compression efficiency and error-propagation range by selecting the segment length $g$ without retraining.
    \item Experiments using real-world network traces demonstrate that ReLViC provides stable reconstruction over a wide range of loss rates and outperforms H.265+FEC and GRACE under severe packet loss.
\end{itemize}

The remainder of this paper is organized as follows. Section~\ref{sec:related_work} reviews related work on video coding and loss resilience. Section~\ref{sec:method} presents ReLViC. Section~\ref{sec:experiments} describes the experimental setup and discusses the results. Section~\ref{sec:conclusion} concludes the paper.

\section{Related Work}\label{sec:related_work}

This section reviews learned video compression and loss-resilient video transmission, which are two strands of literature most relevant to ReLViC. We first examine the dependency structures that enable efficient neural coding, and then discuss conventional and learning-based mechanisms for packet-loss protection and recovery.

\subsection{Learned Video Compression}
Learned video compression replaces hand-crafted coding modules with trainable transforms, temporal prediction, and entropy models optimized under a rate--distortion objective. The end-to-end low-latency codec in~\cite{Rippel_2019_ICCV}~jointly compresses the signals required for learned compensation and maintains a learned temporal state rather than relying exclusively on conventional motion-compensated reference frames. The DCVC family provides a representative evolution of conditional learned video coding. The original DCVC replaces explicit pixel-domain residual coding with feature-domain conditional coding, allowing learned temporal context to assist representation learning, reconstruction, and entropy modeling~\cite{RN2940}. DCVC-TCM subsequently propagates intermediate features in addition to reconstructed frames and extracts multi-scale temporal contexts for different coding modules~\cite{Sheng2023TCM}. In parallel, DCVC-HEM introduces a hybrid spatial--temporal entropy model that combines a temporal latent prior with parallel-friendly spatial priors and content-adaptive quantization~\cite{Li2022HEM}. These developments progressively move temporal information from an external prediction signal into the learned representation and probability model.

DCVC-DC further increases context diversity in both temporal and spatial dimensions through hierarchical quality patterns, group-based offset diversity, and quadtree-partitioned spatial contexts~\cite{Li_2023_CVPR}. The companion image codec EVC addresses a different aspect of practicality by using mask decay to construct efficient single-model variable-rate codecs~\cite{Wang2023EVC}. DCVC-FM then uses feature modulation to support a wide quality range and periodically refreshes temporal features to stabilize long prediction chains~\cite{10655044}. Finally, DCVC-RT targets deployment efficiency through implicit temporal modeling, a single low-resolution latent representation, cross-device integerization, and module-bank-based rate control~\cite{Jia_2025_CVPR}.

Overall, learned video compression has advanced from end-to-end residual coding toward increasingly context-rich, variable-rate, and real-time conditional codecs. However, these models primarily optimize rate--distortion performance and computational efficiency for intact coded streams. Their spatial, temporal, and entropy-model dependencies are not explicitly designed around packet erasures. 

\begin{figure*}[t]
    \centering
    \includegraphics[width=\linewidth]{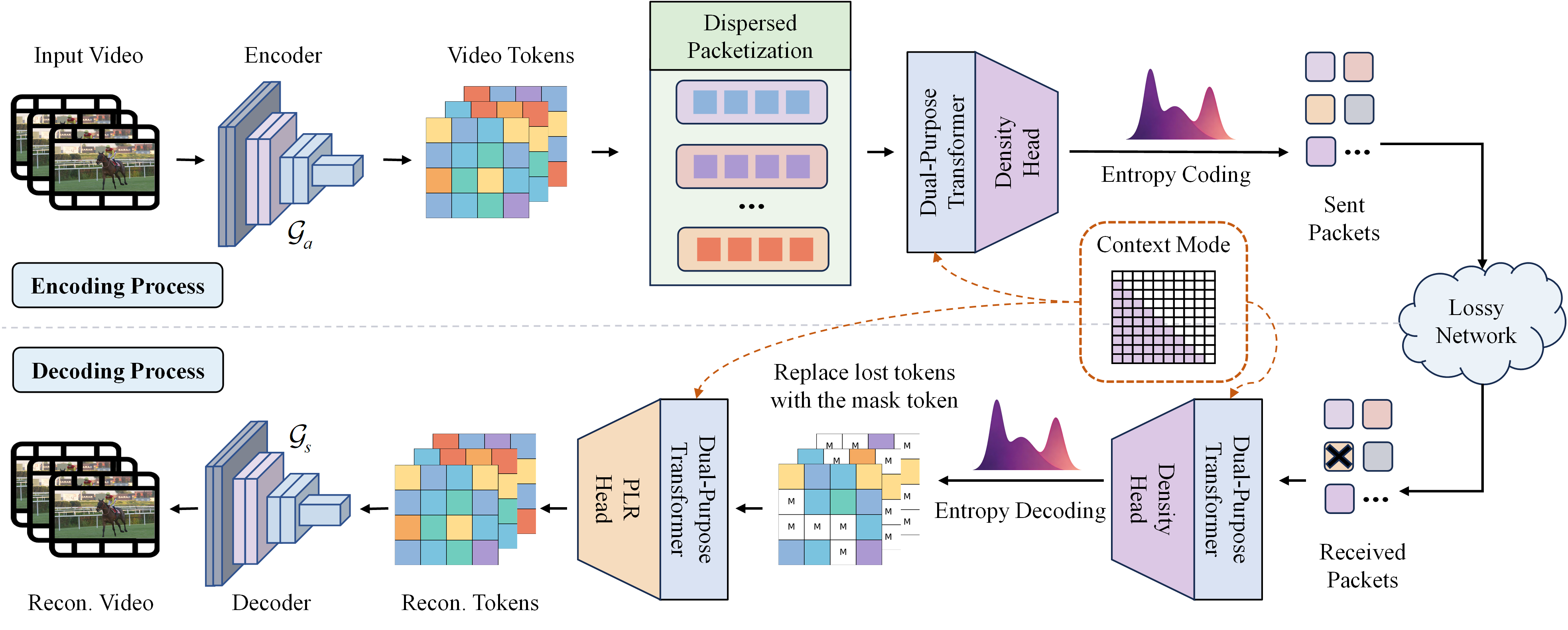}
\caption{Overview of ReLViC.}
    \label{fig:sys_framework}
\end{figure*}

\subsection{Loss-Resilient Video Transmission}

Conventional loss resilience is commonly provided through retransmission and channel coding. In delivery using the Real-time Transport Protocol (RTP), a receiver can report missing packets and the sender can transmit them in a separate retransmission stream~\cite{RFC4588}. ARQ consequently avoids proactive redundancy when no loss occurs, whereas HARQ combines retransmission with error-correcting codes and receiver-side combining~\cite{Ahmed2021HARQ}. Both mechanisms depend on feedback and sufficient time before the playback deadline. Recent production systems accordingly coordinate loss detection, retransmission control, and opportunistic FEC rather than treating them as independent functions~\cite{Wang2026LADR}.

FEC generates repair packets from groups of source packets so that a receiver can recover losses without waiting for a retransmission round trip. RTP FEC supports different protection lengths, group sizes, and unequal protection levels to accommodate media importance and channel conditions~\cite{RFC5109}. Tambur~\cite{Rudow2023Tambur} adapts this principle to video conferencing to improve recovery from burst losses under strict decoding deadlines. The benefit is immediate recovery, but redundancy is transmitted before the realized loss pattern is known, and the protection strength must be selected against a time-varying channel.

Source-coding tools provide complementary protection. Loss-aware packetization and interleaving disperse spatially adjacent coding units across packets, converting a burst loss into separated damaged regions that are easier to conceal~\cite{Hadizadeh2011Burst}. Intra refresh periodically codes selected regions without temporal prediction, thereby limiting error propagation at the cost of additional rate~\cite{Chen2015Adaptive}. When recovery information is unavailable, decoder-side error concealment estimates missing content from received spatial or temporal evidence. For example, neural concealment has been incorporated into VVC decoding to reconstruct missing frames from previously decoded frames~\cite{Benjak2021VVC}. Error concealment can reconstruct an approximate replacement for missing content and mitigate downstream distortion, but it does not recover the lost syntax elements or guarantee encoder--decoder reference consistency.

Learning-based transmission systems increasingly co-design the transmitted representation and its reconstruction mechanism for specific channel conditions. DeepWiVe jointly learns video source coding and wireless channel coding by mapping video directly to channel symbols~\cite{RN2456}, while Gemino uses reference-conditioned neural reconstruction to sustain video conferencing at extremely low bitrates~\cite{Sivaraman2024Gemino}. A recent resilient video tokenizer further learns to reconstruct incomplete token sequences and adapts source--channel coding to channel conditions~\cite{RN3030}. These approaches either target analog joint source--channel transmission or are limited to specific scenarios.

\begin{figure*}[t]
    \centering
    \includegraphics[width=\linewidth]{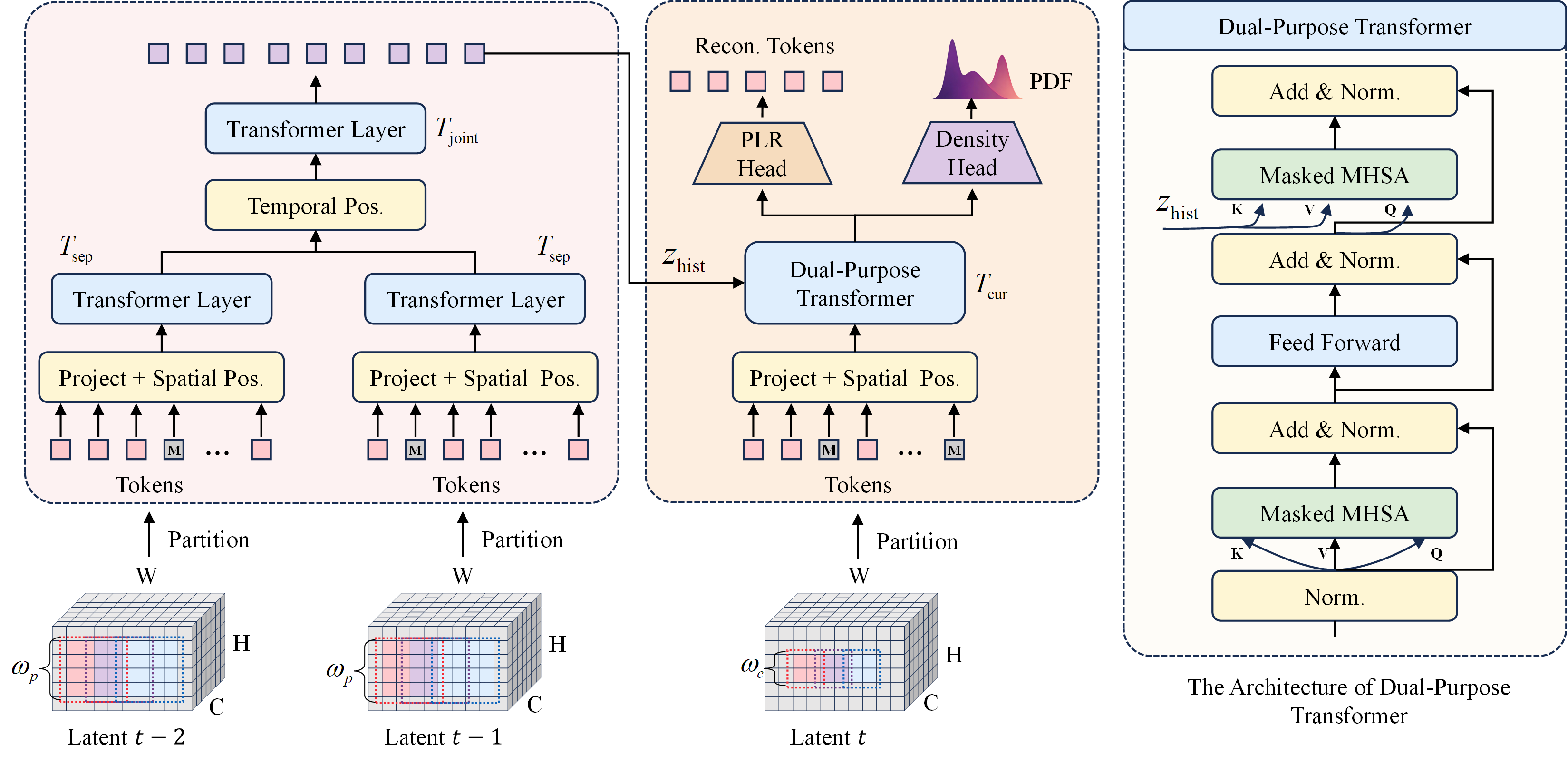}
    \caption{Temporal entropy coding and latent recovery model.}
    \label{fig:entropy_model}
\end{figure*}

\section{Proposed ReLViC Framework}\label{sec:method}
ReLViC is motivated by two effects of packet loss in learned video coding: spatially concentrated missing latents and dependency-driven error propagation. It addresses them by dispersing adjacent tokens across packets and periodically resetting cross-packet context, preserving reconstruction evidence while balancing compression efficiency against propagation range. A shared spatiotemporal Transformer supports both entropy modeling and missing-token recovery, while receiver feedback synchronizes the encoder and decoder states after loss. These considerations motivate the architecture described below.
\subsection{System Overview}
\label{sec:framework}
Fig.~\ref{fig:sys_framework} illustrates ReLViC, which comprises a learned video codec, a token partitioner, a dispersed packetization module, a controllable packet-dependency mechanism, and a dual-purpose Transformer. The framework further employs receiver feedback to synchronize the temporal reference states of the encoder and decoder under packet loss.
\begin{itemize}
    \item \textbf{Learned video codec:} At time $t$, the analysis transform maps an input frame $\boldsymbol{x}_t \in \mathbb{R}^{H \times W \times 3}$ to a continuous latent map $\boldsymbol{y}_t = \mathcal{G}_a(\boldsymbol{x}_t, \boldsymbol{\theta_a}) \in \mathbb{R}^{h \times w \times C}$, where $\mathcal{G}_a(\cdot)$ denotes the encoder that performs the analysis transformation, and $\boldsymbol{\theta_a}$ denotes its parameters, and $C$ is the number of latent channels. This nonlinear transform follows the design principles of ELIC~\cite{9879846}. Strided convolutions reduce the spatial resolution by a factor of $16$, yielding $h=H/16$ and $w=W/16$. Uniform scalar quantization produces the quantized latent map $\boldsymbol{\hat{y}}_t=Q(\boldsymbol{y}_t)$, which is subsequently entropy-coded. The synthesis transform reconstructs the frame as $\boldsymbol{\hat{x}}_t=\mathcal{G}_s(\boldsymbol{\hat{y}}_t,\boldsymbol{\theta_s})$, where $\mathcal{G}_s(\cdot)$ is the decoder that performs the synthesis transformation, and $\boldsymbol{\theta_s}$ denotes its parameters.
    \item \textbf{Token partitioning:} We employ a Transformer to estimate latent-token distributions. Because a Transformer operates on sequential inputs, directly computing attention over an entire video frame would incur prohibitive computational cost. Inspired by conventional block-based video coding, the partitioner reorganizes the quantized latent map and applies a sliding-window mechanism to shorten the input sequence. Given $\boldsymbol{\hat{y}}_t\in \mathbb{R}^{h \times w \times C}$, the partitioner divides it into $\frac{h}{P} \times \frac{w}{P}$ nonoverlapping $P \times P$ windows using $\boldsymbol{\hat{y}}_t^{\mathrm{patch}} = \mathrm{Patch}(\boldsymbol{\hat{y}}_t,P) \in \mathbb{R}^{B\times L\times C}$, where $B=n_h\times n_w$, $L=P^2$, $n_h=\frac{h}{P}$, $n_w=\frac{w}{P}$, and $\mathrm{Patch}(\cdot)$ denotes the partition operation. Each spatial position within a window corresponds to a $C$-dimensional latent token, so each window comprises $L$ latent tokens. A linear projection embeds these tokens into the Transformer dimension as $\boldsymbol{\hat{y}}_t^{\mathrm{emb}}=\boldsymbol{\hat{y}}_t^{\mathrm{patch}}\boldsymbol{W}_{\mathrm{emb}} \in \mathbb{R}^{B\times L\times D}$, where $\boldsymbol{W}_{\mathrm{emb}}$ is a learnable projection matrix, and $D$ denotes the Transformer dimension. This partitioning restricts attention computation to the window-level sequence length, thereby substantially reducing computational complexity.
    \item \textbf{Dispersed packetization:} Under conventional segment-based partitioning, packet loss produces large contiguous regions of missing latent tokens that are difficult to reconstruct. The dispersed packetizer assigns spatially adjacent latent tokens to different packets whenever possible. Consequently, the missing token positions after a packet loss are sparsely distributed over the two-dimensional latent map, enabling the dual-purpose Transformer to reconstruct them from surrounding received tokens. The latent tokens are divided into $K$ network packets.
    \item \textbf{Controllable packet dependencies:} Interpacket dependencies determine the context available for entropy decoding. Independent packets prevent cross-packet error propagation but cannot exploit cross-packet redundancy. Using previously decoded packets as references improves compression efficiency, but the loss of a reference packet can affect subsequent dependent tokens. ReLViC uses a segment-length parameter $g$ and a matrix $\boldsymbol{G}_g\in\{0,1\}^{K\times K}$, which depends on $g$ as we shall see in more details later, to define a periodic-reset finite-segment dependency topology. Selecting $g$ controls the maximum error-propagation span without retraining.
    \item \textbf{Dual-purpose Transformer:} The shared Transformer $T_\mathrm{cur}$ is the core module for predicting latent-token probability distributions and reconstructing missing latent tokens. It uses the historical context representation $\boldsymbol{z}_{\mathrm{hist}}$ derived from the preceding $N$ latent maps when processing the current latent map. By jointly encoding these $N$ latent maps, $\boldsymbol{z}_{\mathrm{hist}}$ summarizes the complete historical window. Through self-attention over the current latent tokens and cross-attention to $\boldsymbol{z}_{\mathrm{hist}}$, $T_\mathrm{cur}$ either predicts probability distributions or reconstructs missing latent tokens. During training, a random mask matrix $\boldsymbol{M} \in \{0, 1\}^{B\times L}$ simulates token loss, where $\boldsymbol{M}_{ij} = 1$ indicates that the $(i,j)$-th latent token is masked. Each masked token is replaced by a learned vector $\boldsymbol{e}\in\mathbb{R}^{C}$.
    \item \textbf{Packet-loss detection and state synchronization:} In a practical deployment, the receiver identifies lost or corrupted packets using packet identifiers and cyclic redundancy checks. After the decoding deadline of frame $t$, the receiver sends a reliable, frame-indexed loss bitmap to the sender. The encoder and decoder then update their temporal reference states according to the same packet-loss pattern before processing frame $t+1$. This feedback-assisted update prevents the temporal contexts maintained at the two endpoints from diverging after packet loss.
\end{itemize}

\subsection{Spatiotemporal Context Modeling for Latent Tokens}
\label{sec:context_model}
Fig.~\ref{fig:entropy_model} illustrates the extraction of the historical context representation and the architecture and processing pipeline of $T_\mathrm{cur}$.
\subsubsection{Asymmetric Tokenization}
We employ asymmetric tokenization to provide a broad historical receptive field. The two preceding latent maps, $\boldsymbol{\hat{y}}_{t-2}$ and $\boldsymbol{\hat{y}}_{t-1}$, are partitioned into $\omega_p\times\omega_p$ windows, whereas the current latent map $\boldsymbol{\hat{y}}_t$ is partitioned into smaller $\omega_c\times\omega_c$ windows. The current-frame windows are nonoverlapping, while the historical windows overlap with a stride of $\omega_c$ because $\omega_p>\omega_c$. This design incorporates a larger neighborhood of textures and structures into each current-frame window, providing richer context for probability-distribution prediction. For clarity, Fig.~\ref{fig:entropy_model} illustrates $\omega_p=4$ and $\omega_c=2$, whereas the implementation uses $\omega_p=8$ and $\omega_c=4$.

\subsubsection{Joint Context Encoder}
The two preceding frames provide the temporal context, i.e., $N=2$. Tokenization and projection yield $\boldsymbol{\hat{y}}_{t-2}^{\mathrm{emb}}$ and $\boldsymbol{\hat{y}}_{t-1}^{\mathrm{emb}}$, to which spatial positional embeddings are added. The spatial Transformer $T_{\mathrm{sep}}$ then processes each frame independently. Its outputs are concatenated along the sequence dimension, augmented with temporal positional embeddings, and passed to $T_\mathrm{joint}$. The resulting $\boldsymbol{z}_{\mathrm{hist}}$ is the historical context representation that summarizes the joint spatiotemporal context of the two preceding frames.

\subsubsection{Dual-Purpose Transformer}
$T_\mathrm{cur}$ estimates latent-token distributions and reconstructs missing latent tokens. The current latent map is tokenized as $\boldsymbol{\hat{y}}_t^{\mathrm{patch}}$ and projected to $\boldsymbol{\hat{y}}_t^{\mathrm{emb}}$, after which positional embeddings are added. Multihead self-attention and cross-attention jointly process the current latent tokens and $\boldsymbol{z}_{\mathrm{hist}}$, with $\boldsymbol{z}_{\mathrm{hist}}$ serving as the keys and values in cross-attention. The density head $f_\mathrm{density}$ comprises a multilayer perceptron (MLP) and nonlinear activation functions and outputs the mean and unconstrained scale parameters as
\begin{equation}
    \boldsymbol{\mu}_t, \tilde{\boldsymbol{\sigma}}_t
    = f_\mathrm{density}\left(T_{\mathrm{cur}}\left(\boldsymbol{c}_t, \boldsymbol{z}_{\mathrm{hist}}\right)\right).
\end{equation}
The outputs satisfy $\boldsymbol{\mu}_t,\tilde{\boldsymbol{\sigma}}_t\in\mathbb{R}^{B\times L\times C}$, and $\boldsymbol{c}_t$ denotes the context in $\boldsymbol{\hat{y}}^{\mathrm{patch}}_t$ that is visible to the decoder. A positive scale is obtained as $\boldsymbol{\sigma}_t=\log\left(1+e^{\tilde{\boldsymbol{\sigma}}_t}\right)\in\mathbb{R}_{>0}^{B\times L\times C}$. Assuming conditional independence across channels, the coding cost of the current frame is
\begin{equation}
    R=-\sum_{b=1}^{B}\sum_{i=1}^{L}\sum_{c=1}^{C}
    \log_2 P_{\hat{Y}}\left(\hat{y}^{\mathrm{patch}}_{t,b,i,c}\mid
    \mu_{t,b,i,c},\sigma_{t,b,i,c}\right),
\end{equation}
where $P_{\hat{Y}}(\cdot)$ is the conditional probability mass function of a quantized latent symbol. The entropy model predicts the parameters of a continuous Gaussian distribution, whereas arithmetic coding requires discrete probabilities. We therefore integrate the distribution over the quantization interval. For a quantization step size $\Delta$,
\begin{equation}
    \begin{aligned}
    P_{\hat{Y}}\bigl(
    &\hat{y}^{\mathrm{patch}}_{t,b,i,c}\mid
    \mu_{t,b,i,c},\sigma_{t,b,i,c}\bigr)\\
    &=
    \int_{k-\frac{\Delta}{2}}^{k+\frac{\Delta}{2}}
    \mathcal N\left(u\mid\mu_{t,b,i,c},\sigma_{t,b,i,c}^2\right)\mathrm{d}u \\
    &=
    \Phi\left(\frac{k+\frac{\Delta}{2}-\mu_{t,b,i,c}}{\sigma_{t,b,i,c}}\right)
    -
    \Phi\left(\frac{k-\frac{\Delta}{2}-\mu_{t,b,i,c}}{\sigma_{t,b,i,c}}\right),
    \end{aligned}
\end{equation}
where $k=\hat{y}^{\mathrm{patch}}_{t,b,i,c}$, and $\Phi(\cdot)$ denotes the standard Gaussian cumulative distribution function. The recovery head $f_\mathrm{PLR}$ predicts the missing latent tokens. Although distribution estimation and latent-token recovery use different visibility patterns, both functions rely on the same spatiotemporal evidence. Sharing the backbone for spatiotemporal context extraction substantially reduces the parameter count and computational cost. The recovery head comprises an MLP and nonlinear activation functions and is given by
\begin{equation}
\label{eq:plr_prediction}
    \boldsymbol{\hat y}_t^\mathrm{pred} = f_\mathrm{PLR}\left(T_{\mathrm{cur}}\left(\boldsymbol{c}_t, \boldsymbol{z}_{\mathrm{hist}}\right)\right),
\end{equation}
where $\boldsymbol{\hat y}_t^\mathrm{pred}\in\mathbb{R}^{B\times L\times C}$. The final latent representation obtained via the combination of predicted and received tokens is expressed as
\begin{equation}
\label{eq:latent_recovery}
    \tilde{\boldsymbol{y}}_{t,M}^{\mathrm{patch}}=\boldsymbol{\hat y}_t^\mathrm{pred}\odot\boldsymbol{M} + \boldsymbol{\hat y}_t^{\mathrm{patch}}\odot \left(1-\boldsymbol{M}\right),
\end{equation}
where $\odot$ denotes elementwise multiplication.

\subsection{Three-Stage Progressive Training}
\label{sec:training}
We adopt a three-stage progressive training strategy that extends single-frame latent compression to temporal conditional modeling and subsequently introduces loss resilience. Stage~I establishes single-frame rate--distortion coding by training $\mathcal{G}_a$ and $\mathcal{G}_s$ while freezing the context modules. Stage~II freezes the image codec and trains $T_\mathrm{sep}$, $T_\mathrm{joint}$, and $T_\mathrm{cur}$ to predict the current-frame distribution from historical latent context. Stage~III jointly fine-tunes all modules using random masks.

\subsubsection{Stage I: Single-Frame Compression}
Stage~I excludes temporal modeling and trains the spatial codec backbone, comprising $\mathcal{G}_a$ and $\mathcal{G}_s$, under a single-frame rate--distortion objective. It also produces stable latent representations for subsequent training. The Stage~I loss includes both the conditional main-latent rate and the hyperlatent rate:
\begin{equation}
\label{eq:training_stage1}
\begin{aligned}
    \mathcal{L}_E = \mathbb{E}_{\boldsymbol{x}_t\sim p_{\mathrm{data}}}\bigg[&-\log_2 p\left(\boldsymbol{\hat{y}}^{\mathrm{patch}}_t\mid\boldsymbol{\hat z}_t\right)-\log_2 p\left(\boldsymbol{\hat z}_t\right)\\
    &+\lambda_1 d\left(\boldsymbol{x}_t,\boldsymbol{\hat x}_t\right) \bigg].
\end{aligned}
\end{equation}
The first two terms in the bracket of the right-hand side of \eqref{eq:training_stage1} estimate the main-latent and hyperlatent rates, respectively. $\boldsymbol{\hat z}_t$ denotes the quantized hyperlatent. $\lambda_1$ controls the rate--distortion trade-off. The function $d(\cdot)$ denotes the mean squared error (MSE). This stage optimizes $\boldsymbol{\theta_a}$, $\boldsymbol{\theta_s}$, and the hyperprior parameters. The hyperprior module is used only in Stage~I and is discarded in the subsequent stages.

\subsubsection{Stage II: Temporal Entropy Modeling}
Stage~II freezes the Stage~I codec and trains $T_{\mathrm{sep}}$, $T_{\mathrm{joint}}$, and $T_{\mathrm{cur}}$. The model learns the conditional distributions of current-frame tokens from historical context, thereby reducing the entropy-coding rate. Because no historical frame is available for the first frame, a learnable reference frame is introduced as a virtual reference frame during training. Entropy decoding proceeds autoregressively. We use an upper-triangular causal mask $\boldsymbol{M}_\mathrm{causal}\in\{0,1\}^{L\times L}$, where $[\boldsymbol{M}_{\mathrm{causal}}]_{ij}=1_{\{i<j\}}$. $i$ indexes the query position, $j$ indexes the key position, and a mask value of one prevents position $i$ from attending to position $j$. Therefore, all future positions $j>i$ are masked, whereas the current and preceding positions $j\leq i$ remain visible. Together with the one-position right shift and the learned start token $s_0\in\mathbb{R}^{C}$, this construction ensures that the distribution of token $i$ is predicted using only the previously decoded tokens $1,\ldots,i-1$. The Stage~II loss is
\begin{equation}
    \mathcal{L}_V = -\mathbb{E}_{\boldsymbol{x}_t\sim p_{\mathrm{data}}}\log_2 p\left(\boldsymbol{\hat{y}}_t^{\mathrm{patch}}\right).
\end{equation}
This stage optimizes $\boldsymbol{\theta}_{\mathrm{sep}}$, $\boldsymbol{\theta}_{\mathrm{joint}}$, and $\boldsymbol{\theta}_{\mathrm{cur}}$, the parameters of $T_{\mathrm{sep}}$, $T_{\mathrm{joint}}$, and $T_{\mathrm{cur}}$, respectively.
\subsubsection{Stage III: Packet-Loss Recovery}
Stage~III jointly fine-tunes all parameters through masked latent modeling. At each iteration, the masking ratio is sampled as $r\sim U(0.05,0.99)$. The masked part of $\boldsymbol{\hat{y}}_t^{\mathrm{patch}}$ is replaced by the learned latent vector $\boldsymbol{e}\in\mathbb{R}^{C}$, yielding the masked sequence $\boldsymbol{\hat y}_{t,M}^{\mathrm{patch}}$, which serves as $\boldsymbol{c}_t$ in \eqref{eq:plr_prediction}. Unlike the causal setting in Stage~II, recovery permits bidirectional interactions among all received tokens. Applying the recovery head and combining the predicted and received tokens according to \eqref{eq:latent_recovery} yields the recovered sequence $\tilde{\boldsymbol{y}}_{t,M}^{\mathrm{patch}}$. We use the following two-branch distortion objective to balance reconstruction quality without packet loss against recovery quality:
\begin{equation}
\label{eq:d_total}
    d_{\mathrm{total}} = \frac{d\left(\boldsymbol{x}_t, \boldsymbol{\hat x}_t\right) + \alpha d\left(\boldsymbol{x}_t, \boldsymbol{\hat x}_{t,M}\right)}{1+\alpha},
\end{equation}
where $\boldsymbol{\hat x}_{t,M}=\mathcal{G}_s\left(\boldsymbol{\hat{y}}_{t,M}\right)$ is the reconstruction from recovered tokens. The hyperparameter $\alpha$ weights the recovery objective, with a larger value assigning greater importance to the masked reconstruction. $\boldsymbol{\hat y}_{t,M}$ is computed as
\begin{equation}
\boldsymbol{\hat y}_{t,M}=\mathrm{Unpatch}\left(\tilde{\boldsymbol{y}}_{t,M}^{\mathrm{patch}},n_h,n_w\right),
\end{equation}
where $\mathrm{Unpatch}(\cdot,n_h,n_w)$ reassembles the patched tokens into a latent map of dimensions $h\times w\times C$. The Stage~III loss is
\begin{equation}
\label{eq:train_loss_stage3}
\begin{aligned}
    \mathcal{L}
    = \mathbb{E}_{\boldsymbol{x}_t\sim p_{\mathrm{data}}}\Bigg[
    \sum_{\forall \boldsymbol{M}_{ij}=1}
    &-\log_2 p\left(\boldsymbol{\hat{y}}_{t,i,j}^{\mathrm{patch}}\mid
    \boldsymbol{\hat{y}}_{t,M}^{\mathrm{patch}}\right)\\
    &+ \lambda_3 d_{\mathrm{total}}\Bigg].
\end{aligned}
\end{equation}
The first term in the bracket of the right-hand side of \eqref{eq:train_loss_stage3} represents the rate of the masked tokens, and $\lambda_3$ controls the rate--distortion trade-off. The unmasked tokens need not be explicitly optimized for rate. Instead, their values provide conditional context that reduces the coding rate of the masked tokens. Because the mask locations are randomized at every iteration, each token is selected as a masked token and optimized across different iterations, enabling global joint optimization. This design incorporates compression efficiency and loss resilience into a unified training objective.

\begin{table*}[t]
\centering
\renewcommand{\arraystretch}{1.2}
\caption{Overview of the three-stage progressive training strategy}
\label{tab:three-stage}
\begin{tabular}{@{}cp{2.5cm}p{4.5cm}p{2.5cm}c@{}}
\toprule
Stage & Optimized modules & Loss & Mask type & Context frames \\
\midrule
Stage I & $\mathcal{G}_a$, $\mathcal{G}_s$, hyperprior & $\mathcal{L}_E$ (single-frame rate--distortion) & None & 0 \\
\addlinespace
Stage II & $T_{\mathrm{sep}}$, $T_{\mathrm{joint}}$, $T_{\mathrm{cur}}$ & $\mathcal{L}_V$ (temporal conditional rate optimization) & Causal autoregressive mask & 2 \\
\addlinespace
Stage III & Joint fine-tuning of all modules & $\mathcal{L}$ (masked latent recovery and two-branch distortion) & Random mask & 2\\
\bottomrule
\end{tabular}
\end{table*}

\subsection{Dispersed Latent Packetization and Controllable Packet Dependencies}
\label{sec:packetization}
\begin{figure}[t]
    \centering
    \includegraphics[width=\linewidth]{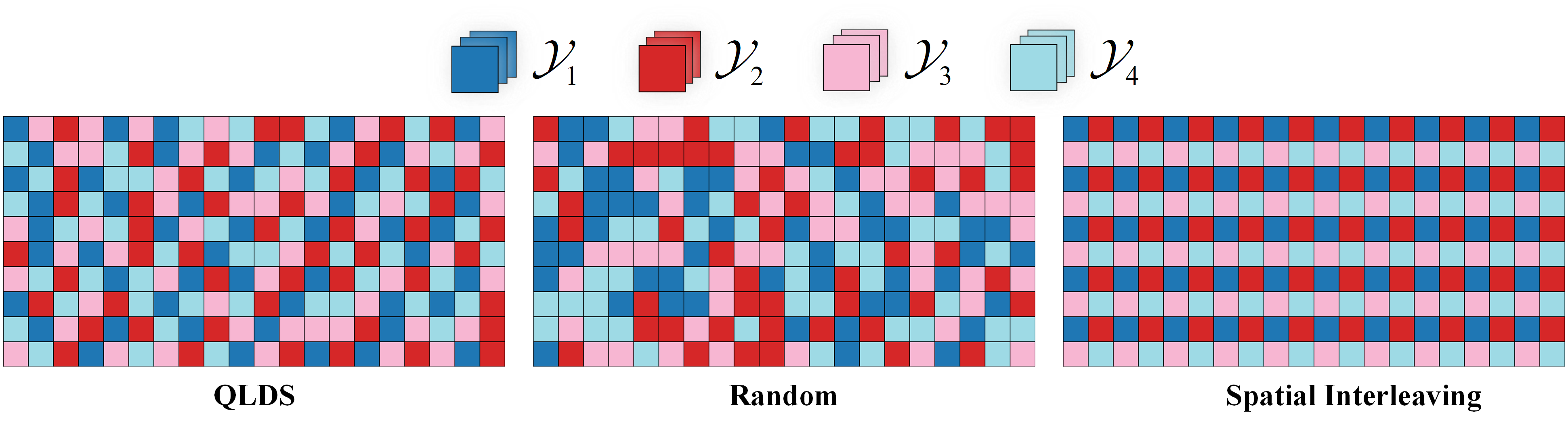}
    \caption{Dispersed latent packetization patterns.}
    \label{fig:pack_mode}
\end{figure}
\subsubsection{Dispersed Latent Packetization}
Dispersed packetization maps the two-dimensional latent tokens to $K$ packets. A contiguous assignment places neighboring tokens in the same packet, so a single packet erasure can produce a large contiguous region of missing tokens. By contrast, a dispersed assignment places nearby tokens in different packets, converting a single erasure into a spatially distributed pattern of missing tokens. The remaining received tokens thus provide local evidence for recovery. This pattern is also consistent with the random masks used during training.

Let $\mathcal{M}_{\mathrm{pack}}(x,y)\in\{0,1,\ldots,K-1\}$ denote the packet index assigned to the token at spatial location $(x,y)$. The token set of packet $k$ is
\begin{equation}
    \mathcal{Y}_k
    =
    \left\{\boldsymbol{\hat y}_t(x,y)\mid \mathcal{M}_{\mathrm{pack}}(x,y)=k\right\}.
\end{equation}
Here, $\boldsymbol{\hat y}_t(x,y)$ denotes the token at location $(x,y)$. Fig.~\ref{fig:pack_mode} illustrates quantized low-discrepancy sequence (QLDS), random, and spatial interleaving assignments.

Random packetization first applies a fixed random permutation $\pi(\cdot)$ to the row-major index $r(x,y)=xw+y$ and then maps the permuted indices to packet identifiers as follows:
\begin{equation}
    \mathcal{M}_{\mathrm{rand}}(x,y)=\pi(r(x,y)) \bmod K.
\end{equation}
Random packetization probabilistically disperses tokens across packets. A sufficiently uniform permutation generally maintains favorable loss-recovery performance. However, the assignment depends on a random seed and may form local clusters. A deterministic mapping provides more controlled spatial coverage.

Spatial interleaving separates neighboring tokens using a regular two-dimensional grid. We factor the packet grid as $K=K_hK_w$ and define the assignment as
\begin{equation}
    \mathcal{M}_{\mathrm{grid}}(x,y)
    =
    \left(x\bmod K_h\right)K_w+\left(y\bmod K_w\right).
\end{equation}
Spatial interleaving provides strictly regular spacing. However, its fixed periodic structure may produce directional or periodic loss patterns when coupled with the latent-map dimensions, packet count, or image structure.
QLDS-based packetization uses a deterministic low-discrepancy sequence to define a two-dimensional traversal order, avoiding both the seed dependence of random assignment and the periodicity of regular interleaving. We therefore use QLDS as the default mapping. Let $\rho$ denote the plastic constant. The sampling coordinates are
\begin{equation}
\label{eq:qlds_packetization}
\left\{
\begin{aligned}
    x_n = \left\lfloor \left( \left( 0.5+n\frac1\rho \right)\bmod1 \right)\cdot h +\frac12 \right\rfloor \bmod h,\\
    y_n = \left\lfloor \left( \left( 0.5+n\frac1{\rho^2} \right)\bmod1 \right)\cdot w +\frac12 \right\rfloor \bmod w.
\end{aligned}
\right.
\end{equation}
$\lfloor\cdot\rfloor$ denotes the floor operator, $n$ is the iteration index, and $h$ and $w$ are the height and width of the latent map, respectively. The location $(x_n,y_n)$ is generated at iteration $n$. The matrix $\mathbf{O}\in\mathbb{R}^{h\times w}$ records the traversal order, with $\mathbf{O}(x,y)$ specifying the order of location $(x,y)$. When the traversal reaches an unvisited location $(x_n,y_n)$, we assign it the current order $m$ by setting $\mathbf O(x_n,y_n)=m$. The normalized order is
\begin{equation}
    \hat{\mathbf{O}}(x,y)=\frac{\mathbf{O}(x,y)}{hw}.
\end{equation}
The packet assignment is then obtained as
\begin{equation}
    \mathcal{M}_{\mathrm{pack}}(x,y)
    =
    \left\lfloor
    \hat{\mathbf{O}}(x,y)\cdot K
    \right\rfloor.
\end{equation}
This mapping spatially disperses the tokens assigned to each packet, thereby reducing the risk of a connected missing region after a single erasure. QLDS is not the only viable layout. Random and spatial-interleaving assignments also perform well when they provide sufficient dispersion. We adopt QLDS because it is deterministic, reproducible, and requires no layout search.
\subsubsection{Packet-Context Modes}
\begin{figure*}[t]
\centering
\subfloat[$g=1$, Intra-slice coding]{\includegraphics[width=0.33\linewidth]{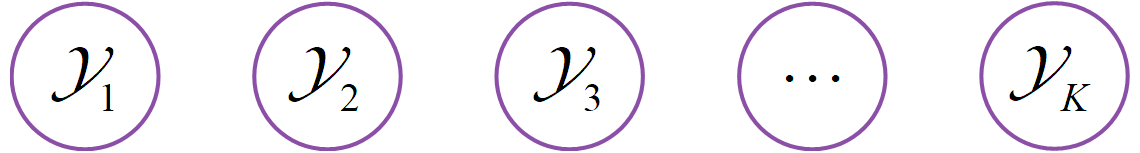}}
\hfil
\subfloat[$g=\frac{K}{2}$, Segmented coding]{\includegraphics[height=1.8cm,keepaspectratio]{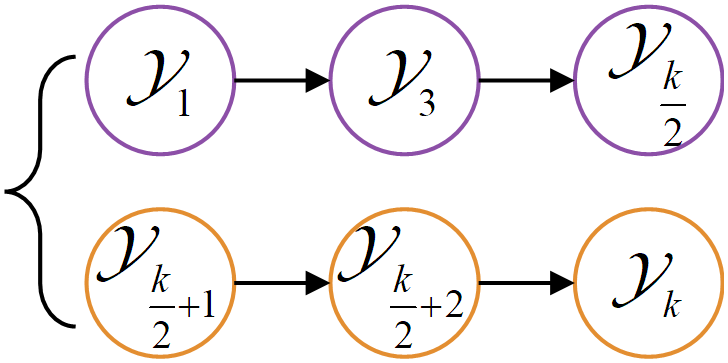}}
\hfil
\subfloat[$g=K$, Layered coding]{\includegraphics[width=0.33\linewidth]{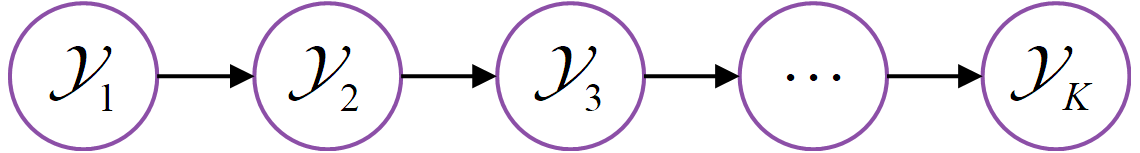}}
\caption{Packet-context modes.}
\label{fig:context_modes}
\end{figure*}

Fig.~\ref{fig:context_modes} illustrates three representative packet-context configurations induced by the segment length $g$. Setting $g=1$ removes interpacket references, an intermediate $g$ periodically resets the reference chain, and $g=K$ allows each packet to use all preceding packets in the frame.

Let $\mathcal{Y}_i$ denote the tokens in packet $i$. The shared Transformer $T_\mathrm{cur}$ estimates their conditional distributions. By the monotonicity of conditional entropy,
\begin{equation}
\begin{aligned}
    H\left(\mathcal{Y}_i\right)
    &\ge H\left(\mathcal{Y}_i\mid \mathcal{Y}_1\right)
    \ge H\left(\mathcal{Y}_i\mid \mathcal{Y}_1,\mathcal{Y}_2\right)\\
    &\ge \cdots
    \ge H\left(\mathcal{Y}_i\mid \mathcal{Y}_1,\ldots,\mathcal{Y}_{i-1}\right).
\end{aligned}
\end{equation}
Additional visible context reduces conditional entropy and can improve compression efficiency, but it also lengthens the paths of error propagation. Inspired by conventional video error-concealment designs~\cite{664283,855913}, we unify the packet-context design by dividing the $K$ packets into finite segments of length $g$. With $m=0,\ldots,\lceil K/g\rceil-1$, define
\begin{equation}
    \mathcal{S}_m=\left\{mg+1,\ldots,\min\left((m+1)g,K\right)\right\}.
\end{equation}
The dependency matrix $\boldsymbol{G}_g\in\{0,1\}^{K\times K}$ is
\begin{equation}
    \boldsymbol{G}_g(i,j)=
    \begin{cases}
    1, & j<i,\quad
    \left\lfloor\dfrac{i-1}{g}\right\rfloor=
    \left\lfloor\dfrac{j-1}{g}\right\rfloor,\\[3pt]
    0, & \text{otherwise}.
    \end{cases}
\end{equation}
Thus, the first packet in every segment is independent, each later packet uses all earlier packets in the same segment, and different segments are completely independent. The matrix is strictly lower triangular, so decoding packet $i$ never uses the current or a future packet. If a packet is lost, only later packets in its segment can inherit the resulting decoding error. Thus, the maximum propagation span is $g-1$ packets.
For the general matrix form, let $\boldsymbol{L}_n$ be the $n\times n$ strictly lower-triangular all-ones matrix. If $K=qg+r$ with $0\le r<g$, then
\begin{equation}
    \boldsymbol{G}_g=\operatorname{blkdiag}\left(
    \underbrace{\boldsymbol{L}_g,\ldots,\boldsymbol{L}_g}_{q\ \text{blocks}},
    \boldsymbol{L}_r\right),
\end{equation}
where the final block $\boldsymbol{L}_r$ is omitted when $r=0$.

\begin{figure}[t]
\centering
\subfloat[$g=1$]{\includegraphics[width=0.4\linewidth]{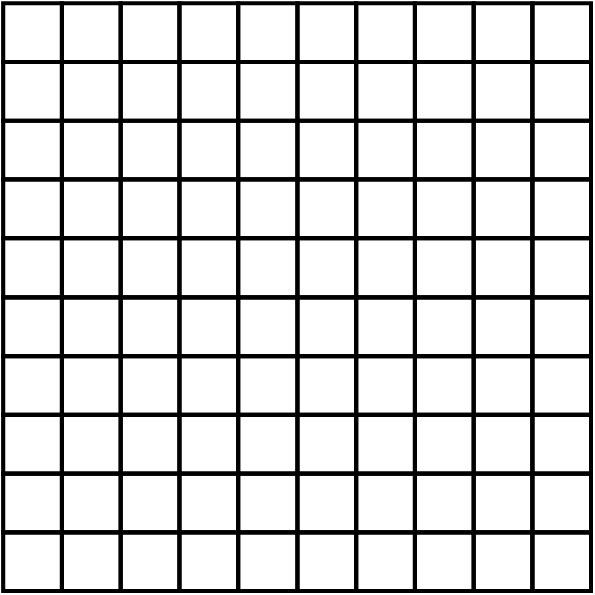}}
\hfil
\subfloat[$g=4$]{\includegraphics[width=0.4\linewidth]{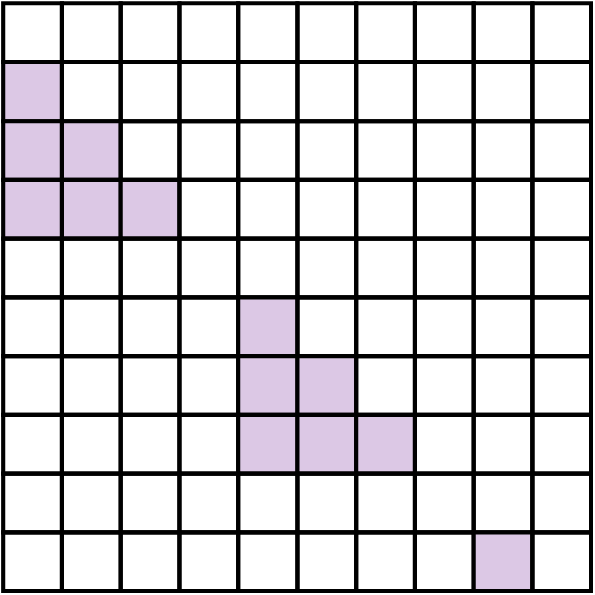}}
\\
\subfloat[$g=7$]{\includegraphics[width=0.4\linewidth]{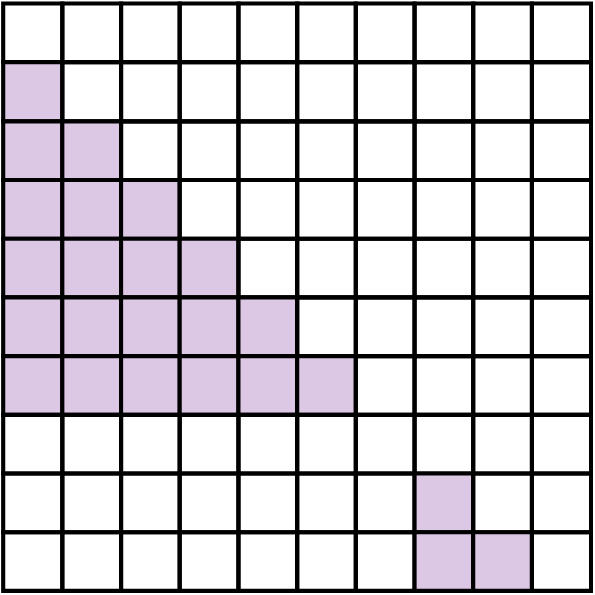}}
\hfil
\subfloat[$g=K$]{\includegraphics[width=0.4\linewidth]{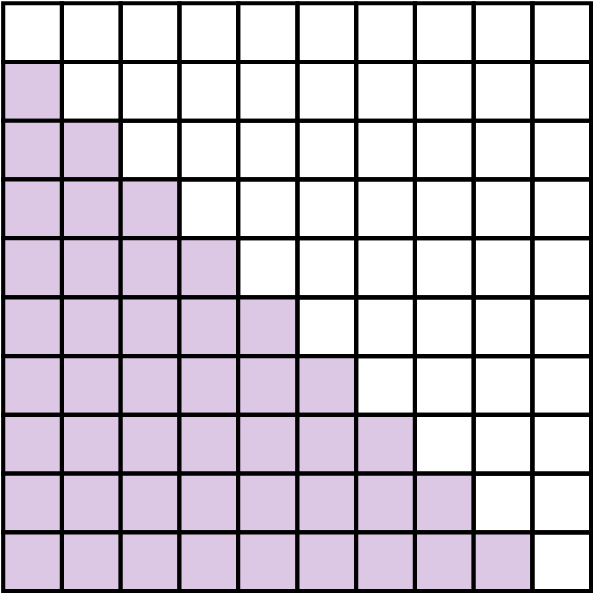}}
\caption{Packet-context dependency matrices.}
\label{fig:context_matrixs}
\end{figure}

Fig.~\ref{fig:context_matrixs} visualizes the resulting dependency matrices for $g=1$, $4$, $7$, and $K$. As $g$ increases, the lower-triangular blocks become larger, providing richer cross-packet context while extending the possible error-propagation path. The loss of any referenced packet propagates along the dependency chain, rendering all subsequent dependent packets undecodable. All packets that are either physically lost or rendered undecodable are uniformly recovered by $f_\mathrm{PLR}$.

\section{Experimental Setup and Results}
\label{sec:experiments}
\subsection{Experimental Setup}
\textbf{Training details:} All three stages are trained on the Vimeo-90k septuplet dataset, which contains 91,701 sequences of seven consecutive frames. Each training input is randomly cropped to $256\times256$. In Stage~I, a $\lambda$ warm-up multiplies $\lambda_1$ in~\eqref{eq:training_stage1} by $10$ during the first $15\%$ of the iterations, prioritizing reconstruction quality while the latent representation is initialized. We set $\lambda_1=0.01$ in Stage~I and use $\lambda_3\in\{0.00125,0.01,0.02,0.04,0.08\}$ in Stage~III. The models are optimized using AdamW. The learning rate is increased linearly to $10^{-4}$ over the first $2\%$ of the iterations and decayed to $10^{-5}$ after $85\%$ of the iterations. Training is performed using two NVIDIA A100 GPUs.

\textbf{Baselines:} We compare ReLViC with H.265 protected by Reed--Solomon (RS) FEC and GRACE~\cite{RN2730}. The notation $\mathrm{H.265}+x\%\mathrm{FEC}$ denotes an H.265 stream with a redundancy ratio of $x=N_r/(N_k+N_r)\times100$, where $N_k$ and $N_r$ are the numbers of source and parity packets, respectively. The RS code recovers the source data upon receiving any $N_k$ of the $N_k+N_r$ transmitted packets. We evaluate $x\in\{0,10,30,50,70\}$. GRACE is an advanced mask-trained learned video codec that masks motion-vector and residual tensors during training to support recovery from packet loss.

\begin{figure}[t]
    \centering
    \includegraphics[width=\linewidth]{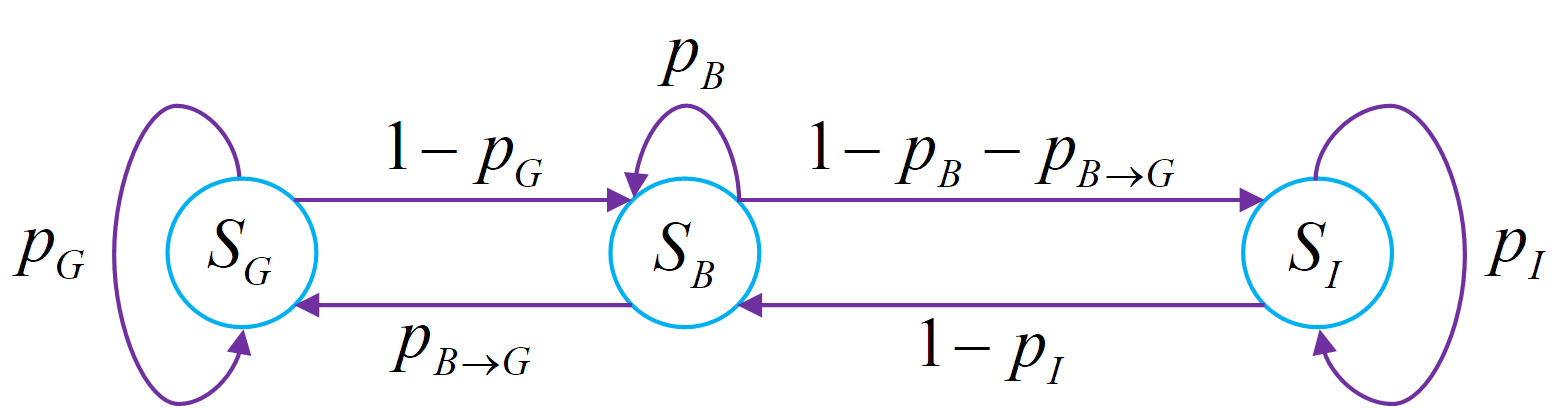}
    \caption{Three-state Markov loss model.}
    \label{fig:markov}
\end{figure}
\begin{figure*}[t]
    \centering
    \includegraphics[width=\linewidth]{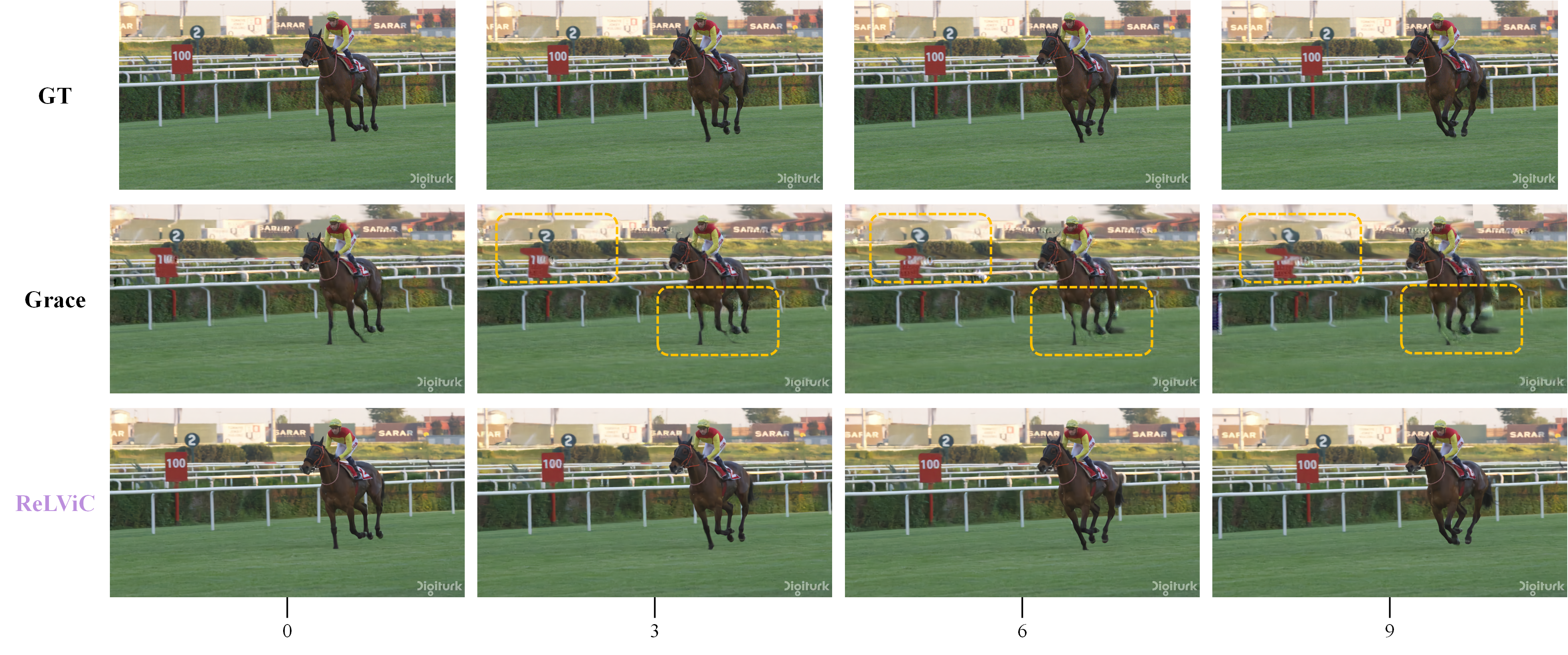}
    \caption{Visual reconstruction comparison at 50\% packet-loss rate.}
    \label{fig:recons_0.5_loss}
\end{figure*}
\textbf{Evaluation details:} We evaluate all methods on the Ultra Video Group (UVG) dataset~\cite{10.1145/3339825.3394937}, which contains seven 1080p videos, and measure reconstruction quality using peak signal-to-noise ratio (PSNR). The plastic constant $\rho$ used in \eqref{eq:qlds_packetization} is 1.324717957~\cite{Iliopoulos31122015}. To balance reconstruction quality and decoding delay, each frame is divided into $10$ packets. The H.265+FEC baseline uses an effective payload budget of 1460 bytes per packet. To improve H.265 decodability, we allocate FEC redundancy non-uniformly among parameter-set network abstraction layer units (VPS/PPS/SPS) and I-frames, P-frames, and B-frames using weights of $6{:}5{:}3{:}1$, respectively. This strategy prioritizes the decodability of critical data types and substantially improves evaluation stability. For frames that fail to decode, we apply a penalty to their PSNR values.

\textbf{Loss simulation:} We simulate burst losses using the three-state Markov model in Fig.~\ref{fig:markov}, which is widely adopted for wireless packet-loss modeling~\cite{749301,1510488}. The states $S_G$, $S_B$, and $S_I$ represent loss-free transmission, packet loss, and intermittent availability, respectively. Packets are always lost in $S_B$ and delivered in $S_G$ and $S_I$. From $S_I$, the link can remain temporarily available or return to $S_B$, modeling intermittent recovery within a loss burst. We use four loss patterns derived from mobile Internet traces~\cite{milner2004analysis}. Table~\ref{tab:markov_loss_params} lists their transition parameters, average loss rate $\epsilon$, and average burst length $\gamma$. The four patterns represent light long bursts, light short bursts, moderate short bursts, and heavy short bursts.

\begin{table}[t]
  \centering
  \renewcommand{\arraystretch}{1.2}
  \caption{Parameters of the three-state Markov loss model}
  \label{tab:markov_loss_params}
  \begin{tabular}{lcccccc}
    \toprule
    Pattern & $p_G$ & $p_B$ & $p_I$ & $p_{B \rightarrow G}$ & $\epsilon$ & $\gamma$ \\
    \midrule
    EP1 & 0.99968 & 0.8462 & 0      & 0.1538 & 0.002 & 6.5  \\
    EP2 & 0.9798  & 0.3372  & 0.3333 & 0.6304 & 0.031 & 1.51 \\
    EP3 & 0.9363  & 0.4072 & 0.5662 & 0.3631 & 0.138 & 1.69 \\
    EP4 & 0.8507  & 0.6305 & 0.2    & 0.2982 & 0.324 & 2.71 \\
    \bottomrule
  \end{tabular}
\end{table}

\subsection{Results}
\textbf{Visual quality:}
Fig.~\ref{fig:recons_0.5_loss} compares ReLViC with GRACE at a 50\% packet-loss rate. ReLViC operates in the intra-slice packet-context mode. Although GRACE preserves recognizable content in the initial frames, distortions and artifacts accumulate along its reference path. In this example, ReLViC retains more visible structure because $T_\mathrm{cur}$ reconstructs missing latent tokens from the received current-frame tokens and temporal context.

\begin{figure*}[t]
    \centering
    \includegraphics[width=\linewidth]{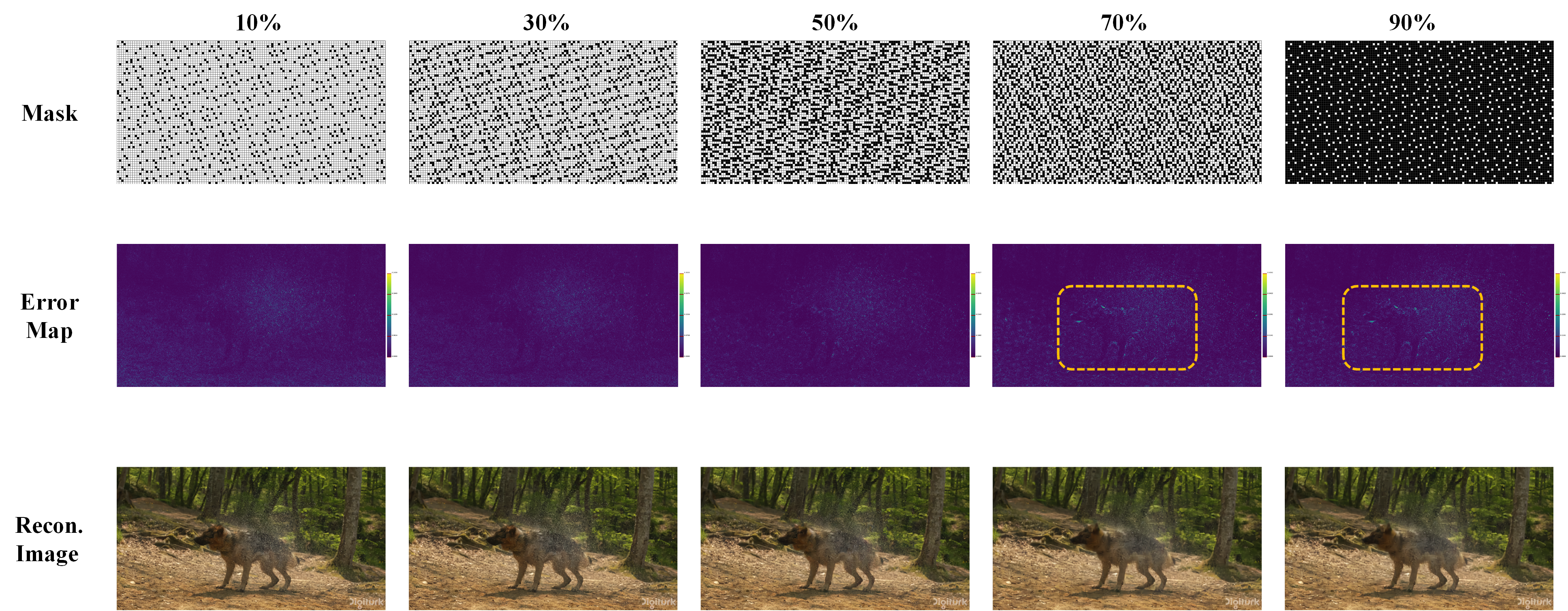}
    \caption{Visual reconstruction comparison across packet-loss rates.}
    \label{fig:recons_vs_loss_rate}
\end{figure*}
\begin{figure*}[t]
    \centering
    \includegraphics[width=\linewidth]{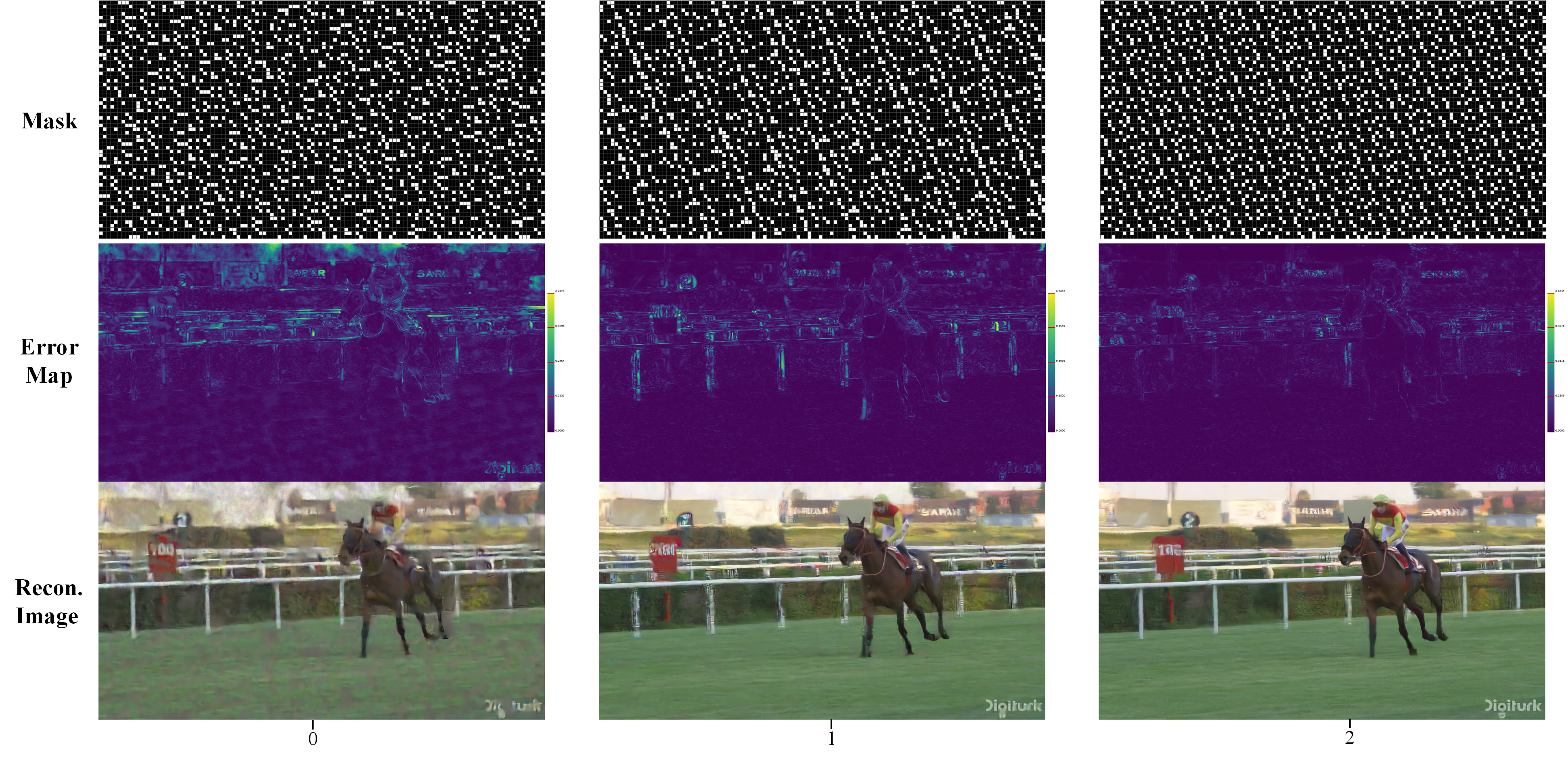}
    \caption{Temporal evolution of reconstruction at 90\% packet-loss rate.}
    \label{fig:recons_0.9_loss}
\end{figure*}
Fig.~\ref{fig:recons_vs_loss_rate} presents reconstructed frames and error maps at different packet-loss rates. The reconstruction error changes only slightly at low loss rates, whereas the degradation becomes more pronounced at loss rates of 70\% and 90\%. The model estimates the missing content from both the received current-frame latents and the historical latent context.

\begin{figure*}[t]
\centering
\subfloat[EP1 (average loss rate 0.2\%, average burst length 6.5)]{\includegraphics[width=0.49\linewidth]{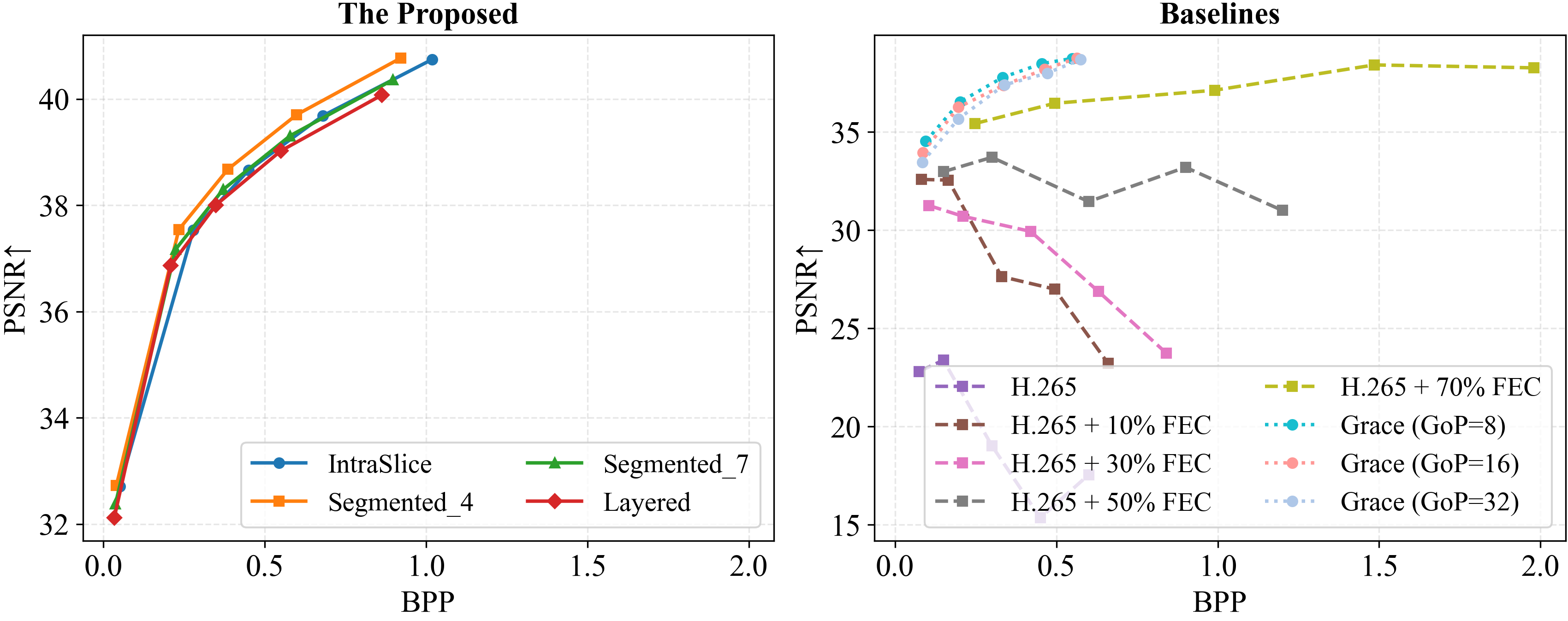}}
\hfil
\subfloat[EP2 (average loss rate 3.1\%, average burst length 1.51)]{\includegraphics[width=0.49\linewidth]{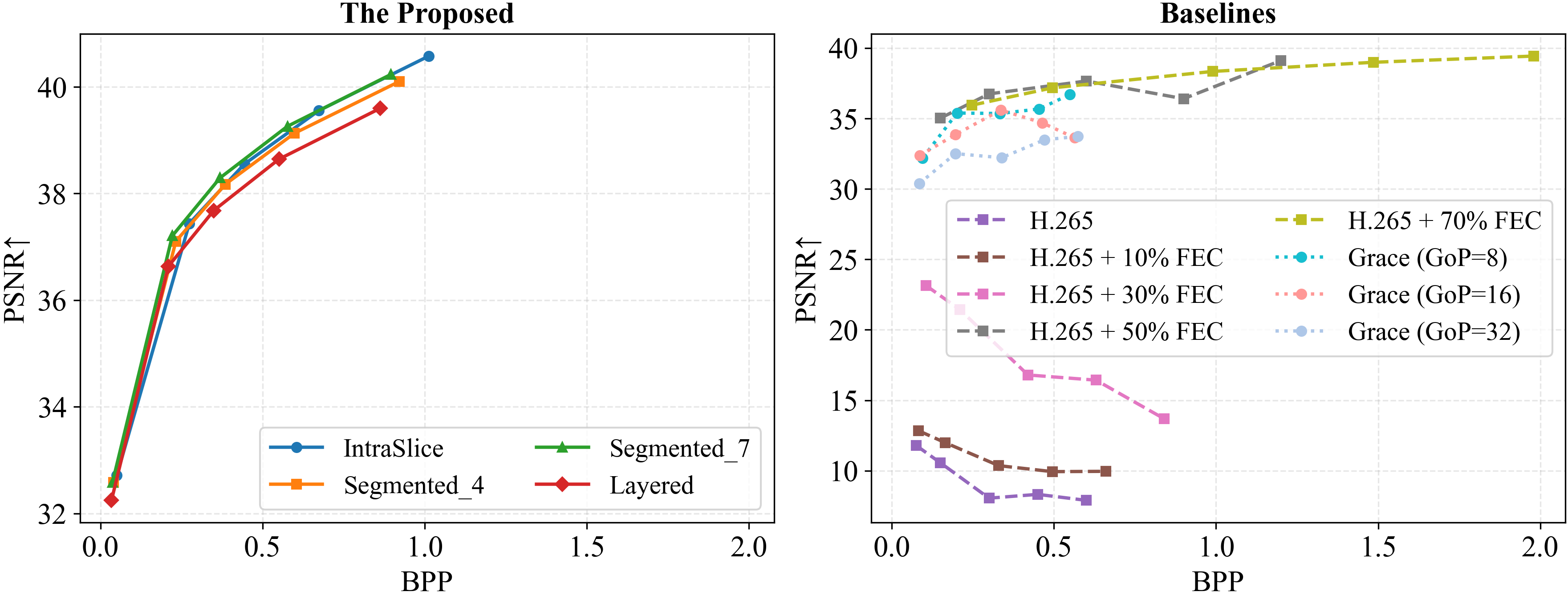}}
\\
\subfloat[EP3 (average loss rate 13.8\%, average burst length 1.69)]{\includegraphics[width=0.49\linewidth]{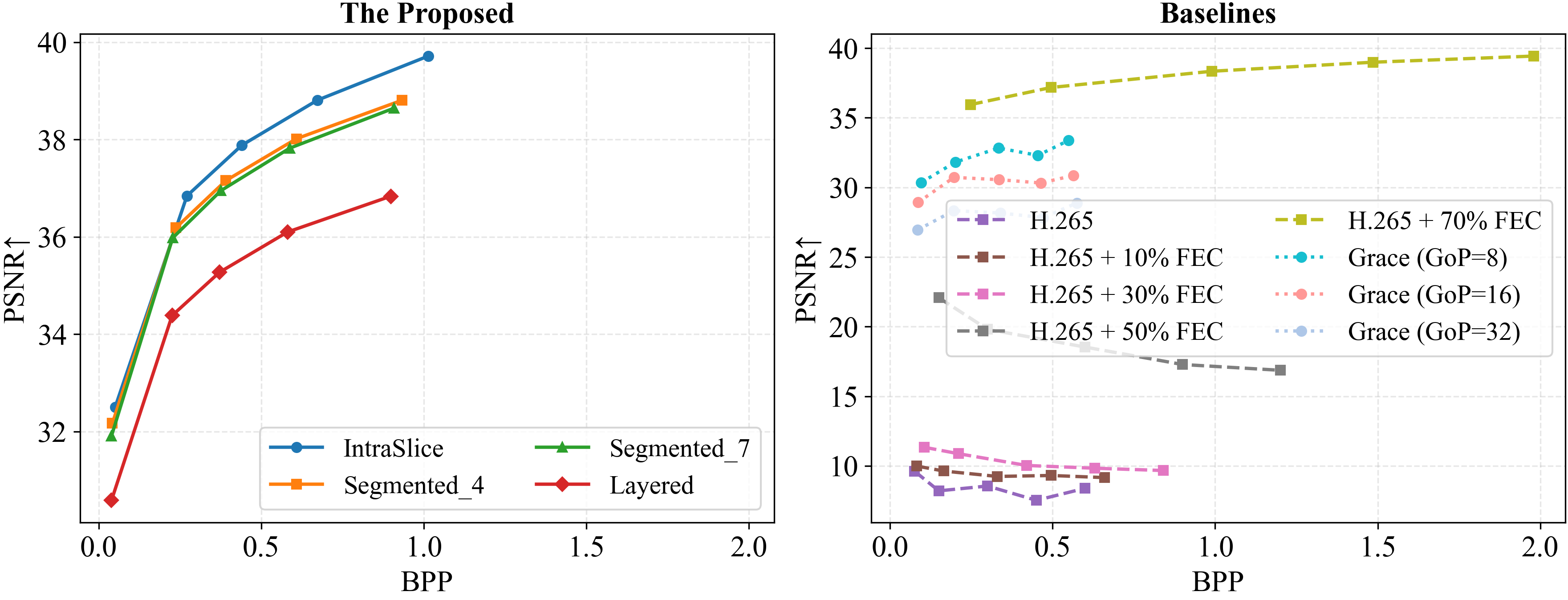}}
\hfil
\subfloat[EP4 (average loss rate 32.4\%, average burst length 2.71)]{\includegraphics[width=0.49\linewidth]{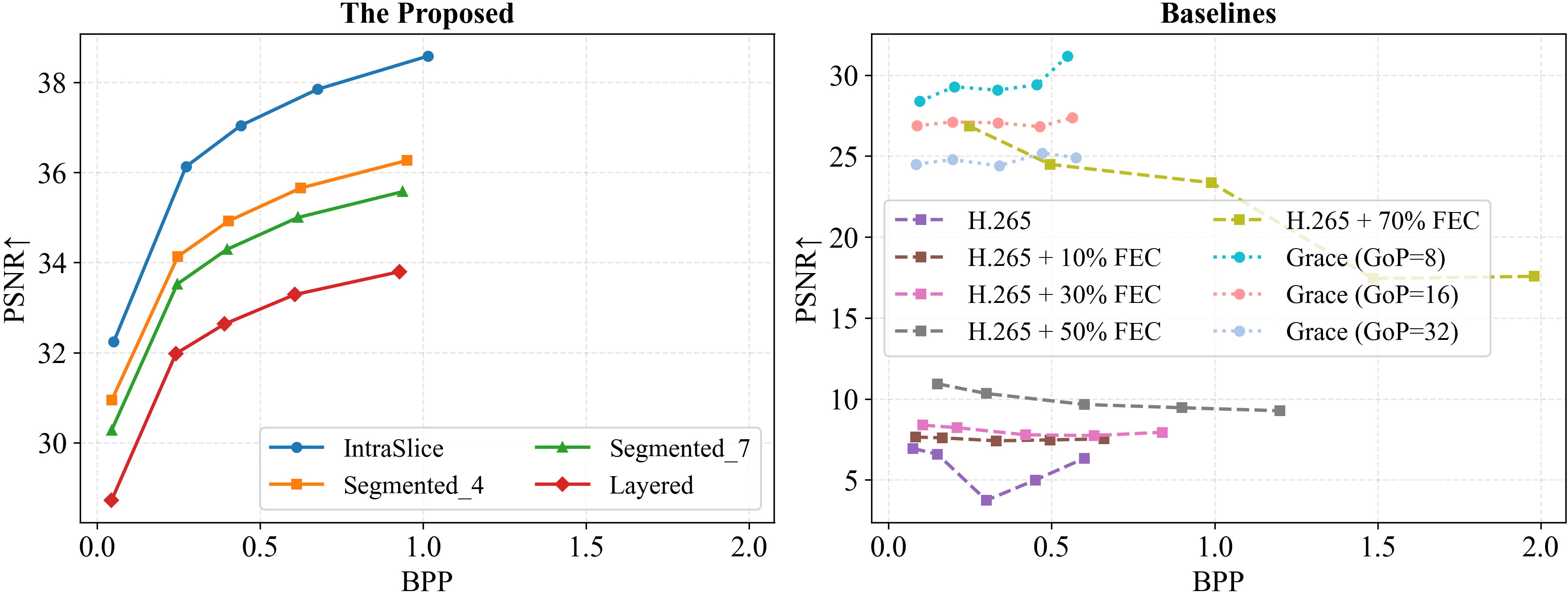}}
\caption{Rate--distortion performance under four burst-loss traces.}
\label{fig:recons_markov_PSNR}
\end{figure*}

\begin{figure}[t]
    \centering
    \includegraphics[width=\linewidth]{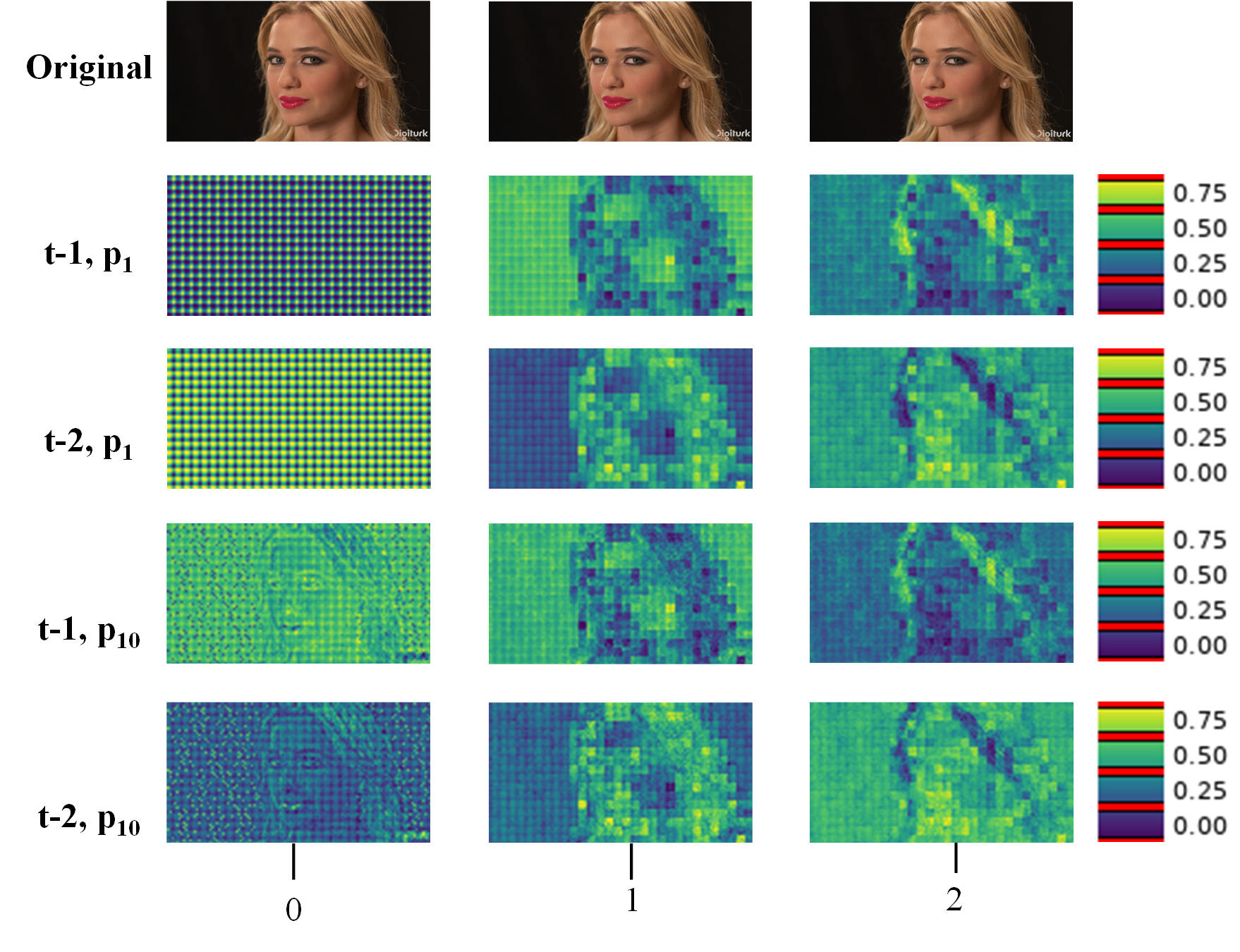}
    \caption{Cross-attention heatmaps of the dual-purpose Transformer.}
    \label{fig:cross_attn}
\end{figure}

Fig.~\ref{fig:recons_0.9_loss} shows the temporal evolution of reconstruction at a 90\% packet-loss rate. Because the first frame has no preceding-frame context, it contains visible artifacts and large regions of smeared color. The next two frames exploit received current-frame tokens and recovered historical context. This additional evidence improves reconstruction and reduces the highlighted regions in the error maps.

Fig.~\ref{fig:cross_attn} visualizes cross-attention to historical frames under the layered packet-context mode. From left to right, the maps correspond to the first three coded frames. The latent maps at $t-1$ and $t-2$ provide temporal context, and $p_N$ denotes the packet currently being encoded. When encoding the first packet of the first frame, no historical frame is available as a reference. Consequently, the cross-attention heatmap exhibits no discernible structure. By the tenth packet of the first frame, previously decoded packets are available as references, and the attention map begins to reveal image-content structure. For subsequent frames, the attention maps indicate that the model identifies motion regions and allocates attention to different spatial areas. Moreover, the attention maps for frames $t-1$ and $t-2$ are complementary, suggesting that the model balances information from the two preceding frames. For example, it assigns more attention to the background in frame $t-1$ and to the face and hair in frame $t-2$.

\textbf{Performance analysis:}
Fig.~\ref{fig:recons_markov_PSNR} compares ReLViC with the baselines under the four loss traces. Although EP1 has a low loss rate, its average burst length is relatively long. The FEC scheme therefore requires substantial redundancy, such as $70\%$, to maintain reasonable performance. Without sufficient redundancy, packet-decoding failures produce severe visual artifacts and sharply reduce PSNR. For GRACE, we evaluate group-of-pictures (GOP) sizes of 8, 16, and 32. A smaller GOP shortens the error-propagation path and consequently improves recovery quality. Across bitrates of approximately 0.1--0.6 bits per pixel (bpp), GRACE performs comparably to ReLViC. Under EP2, the performance of GRACE decreases as the loss probability increases, whereas ReLViC performs better than GRACE and $\mathrm{H.265}+\mathrm{FEC}$. Under EP3 and EP4, which have relatively high loss rates, ReLViC exhibits a clear advantage. Even with $70\%$ redundancy, the FEC scheme experiences numerous decoding errors, while GRACE remains decodable but exhibits markedly lower recovery quality.

EP1 and EP2 have relatively low loss rates. Under these conditions, packet-context modes that use interpacket references, such as layered coding and segmented coding, outperform intra-slice coding, which uses no such references. Under EP3 and EP4, which have high loss rates, intra-slice coding performs better because it does not depend on other packets and therefore avoids interpacket error propagation. The performance differences among the packet-context modes show that selecting the segment length $g$ directly controls the trade-off between compression efficiency and error-propagation range. Switching among these modes therefore enables a flexible balance between compression efficiency and transmission resilience under different network conditions without retraining.

\begin{figure}[t]
    \centering
    \includegraphics[width=\linewidth]{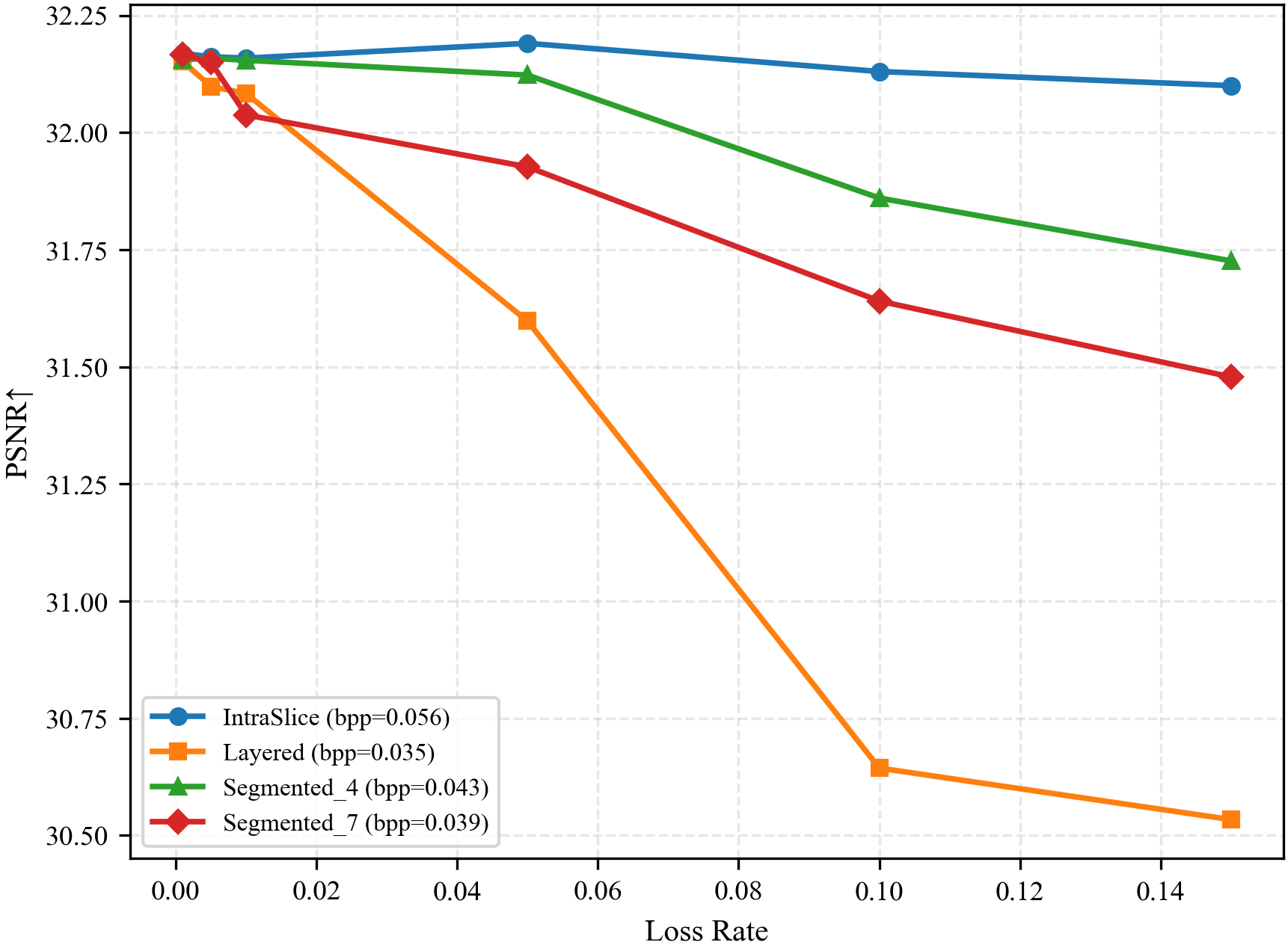}
    \caption{Reconstruction performance of packet-context modes across loss rates.}
    \label{fig:PSNR_vs_LossRate}
\end{figure}

Fig.~\ref{fig:PSNR_vs_LossRate} presents the reconstruction quality of different packet-context modes at different packet-loss rates. Intra-slice coding exhibits greater transmission resilience than the other reference modes. Reference-based coding schemes, such as layered coding, require fewer bits under low packet-loss rates. Overall, as the packet-loss rate increases, coding modes with longer reference chains suffer more rapid performance degradation.

\begin{figure}[t]
    \centering
    \includegraphics[width=\linewidth]{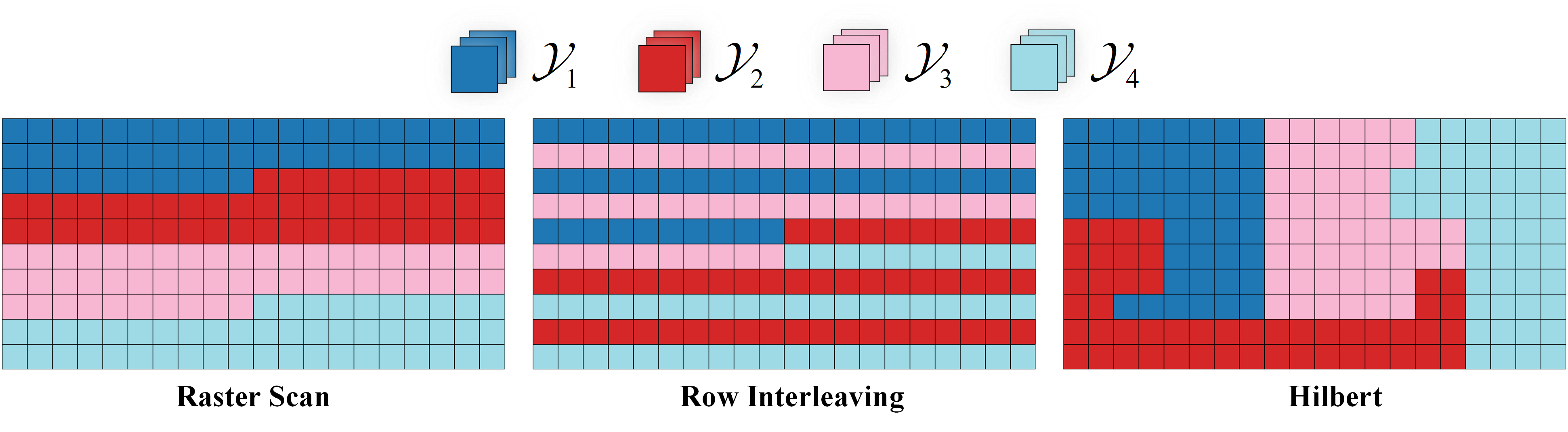}
    \caption{Packetization baselines.}
    \label{fig:pack_baselines}
\end{figure}

\begin{figure}[t]
\centering
\subfloat[$g=1$, Intra-slice coding]{\includegraphics[width=0.49\linewidth]{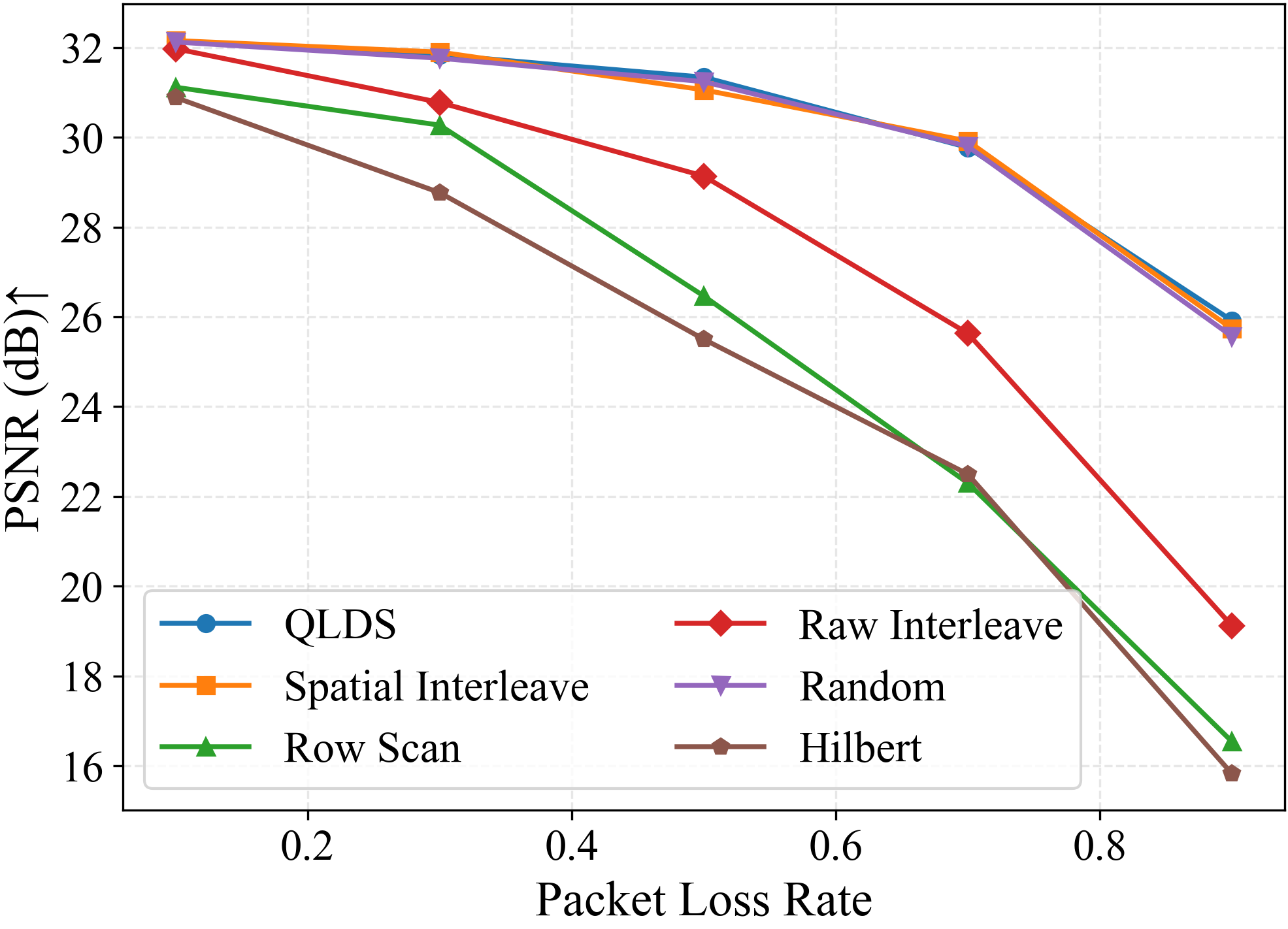}}
\hfil
\subfloat[$g=4$, Segmented coding]{\includegraphics[width=0.49\linewidth]{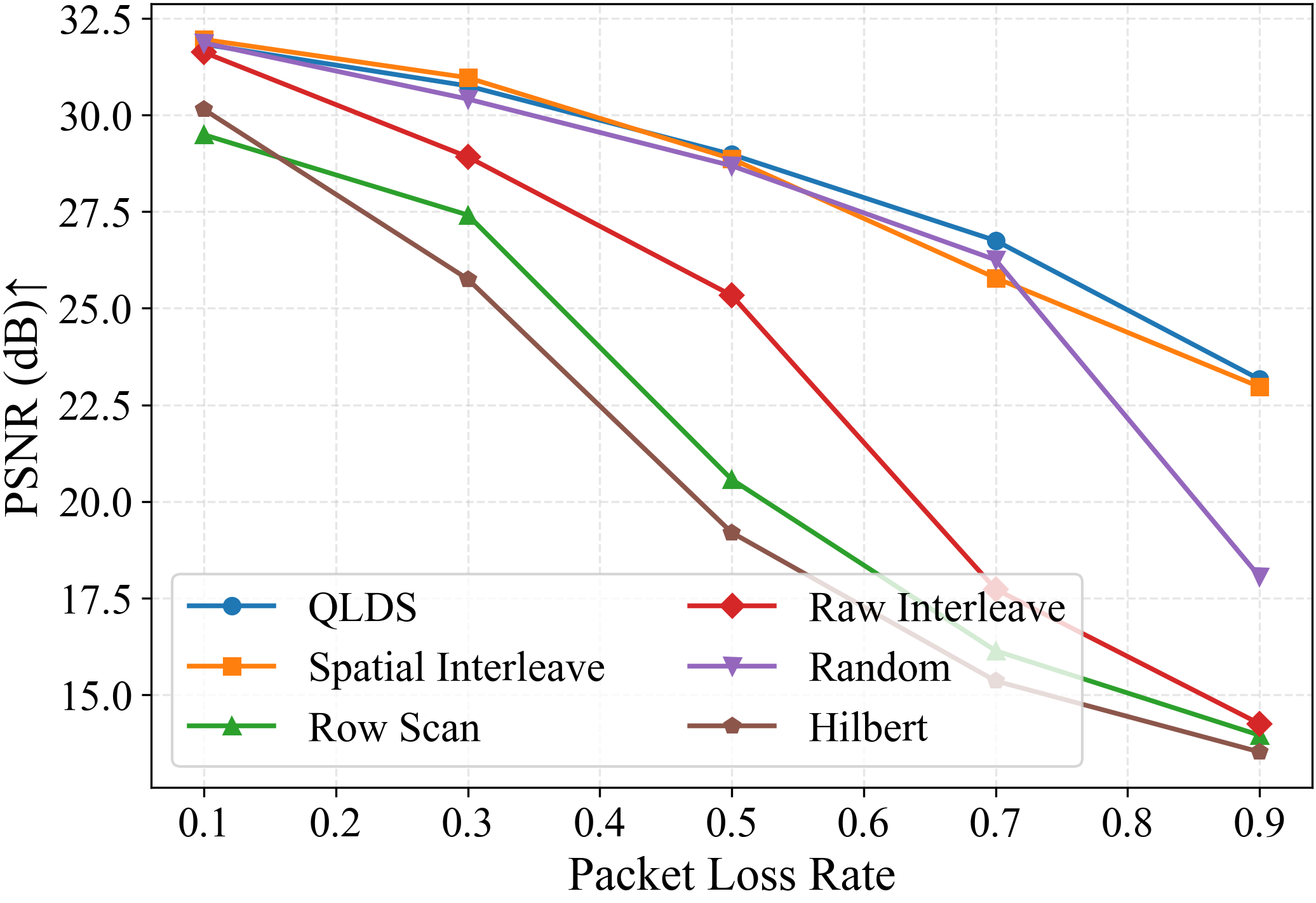}\label{fig:PSNR_vs_pack_mode_b}}
\\
\subfloat[$g=7$, Segmented coding]{\includegraphics[width=0.49\linewidth]{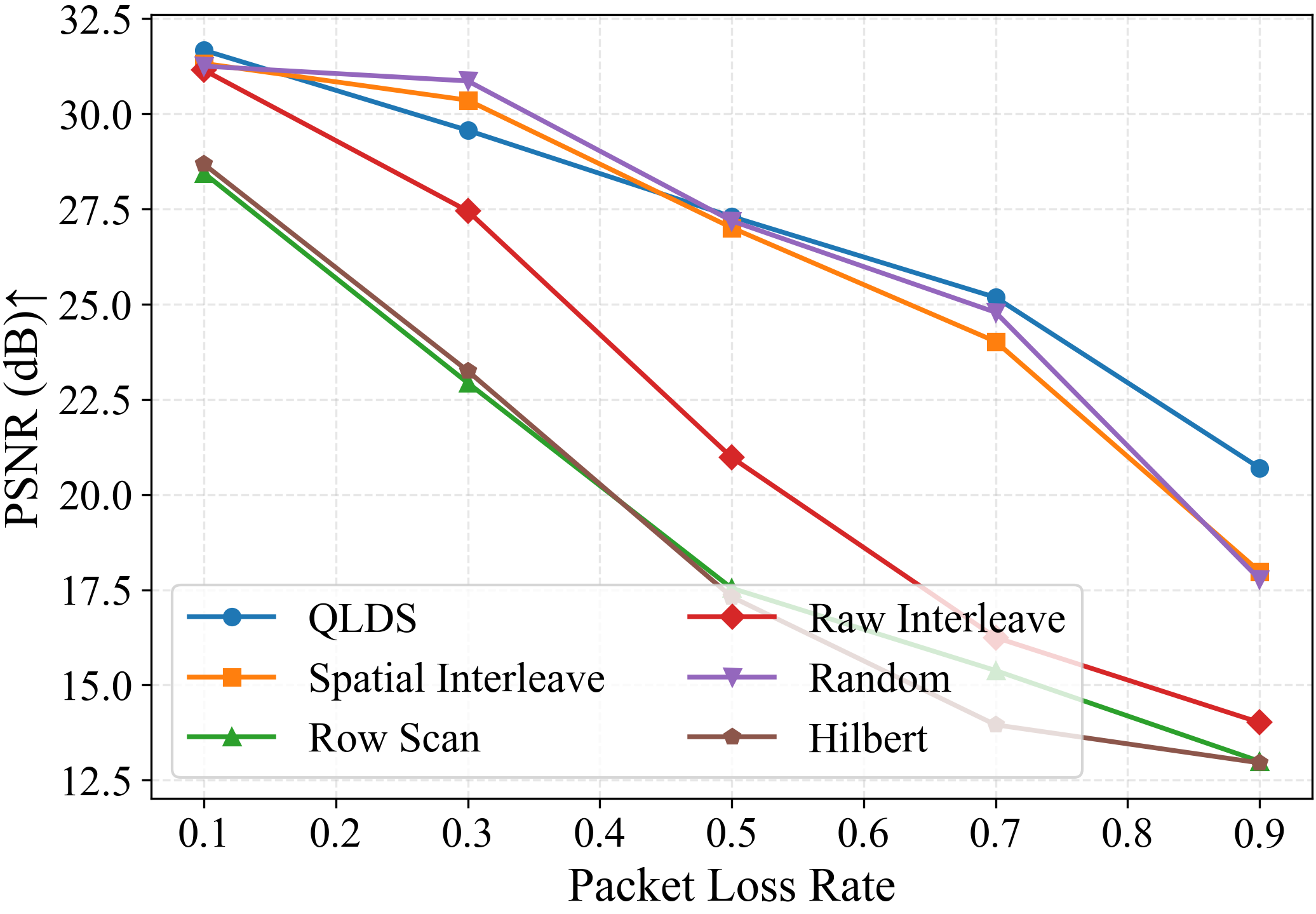}\label{fig:PSNR_vs_pack_mode_c}}
\subfloat[$g=K$, Layered coding]{\includegraphics[width=0.49\linewidth]{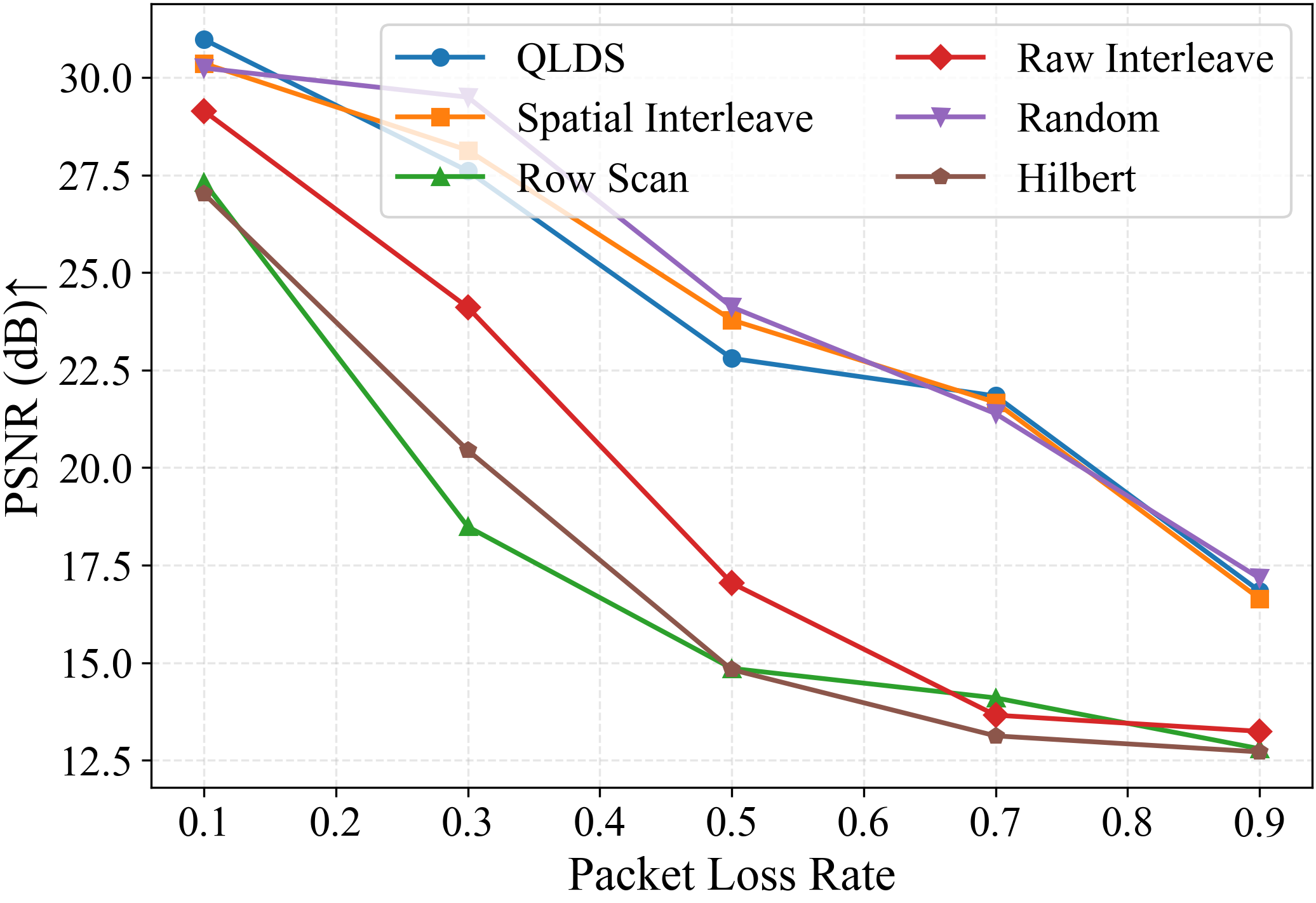}}
\caption{Reconstruction performance of different packetization modes.}
\label{fig:PSNR_vs_pack_mode}
\end{figure}

\textbf{Ablation studies:}
To investigate the effect of spatial dispersion on packet-loss concealment, we compare ReLViC with the three token-packetization baselines shown in Fig.~\ref{fig:pack_baselines}. Raster-scan packetization fills packets contiguously in row-major order and represents a typical contiguous-region packetization method. Row-interleaved packetization assigns packets cyclically along each row, reducing the extent of contiguous missing regions but providing limited two-dimensional dispersion. Hilbert packetization orders tokens according to a Hilbert-curve traversal. Although this approach preserves spatial locality, tokens within the same packet remain strongly clustered. These methods provide contiguous or weakly dispersed baselines for evaluating how spatial clusters of missing tokens affect recovery quality. By comparison, random packetization, spatial-interleaving packetization, and QLDS-based packetization represent probabilistic dispersion, regular-grid dispersion, and deterministic low-discrepancy dispersion, respectively.
\begin{table*}[t]
  \centering
  \renewcommand{\arraystretch}{1.2}
  \setlength{\tabcolsep}{4pt}
  \caption{Model size and computational complexity}
  \label{tab:model_complexity}
  \begin{tabular}{@{}lccccc@{}}
    \toprule
     & Parameters & Encoding complexity & Decoding complexity & Encoding time & Encoding speed \\
    \midrule
    ReLViC & 231.155 M & 32.849 TFLOPs & 32.862 TFLOPs & $\approx 0.8294$ s/frame & $\approx 1.2$ FPS \\
    \bottomrule
  \end{tabular}
\end{table*}

\begin{figure}[t]
    \centering
    \includegraphics[width=\linewidth]{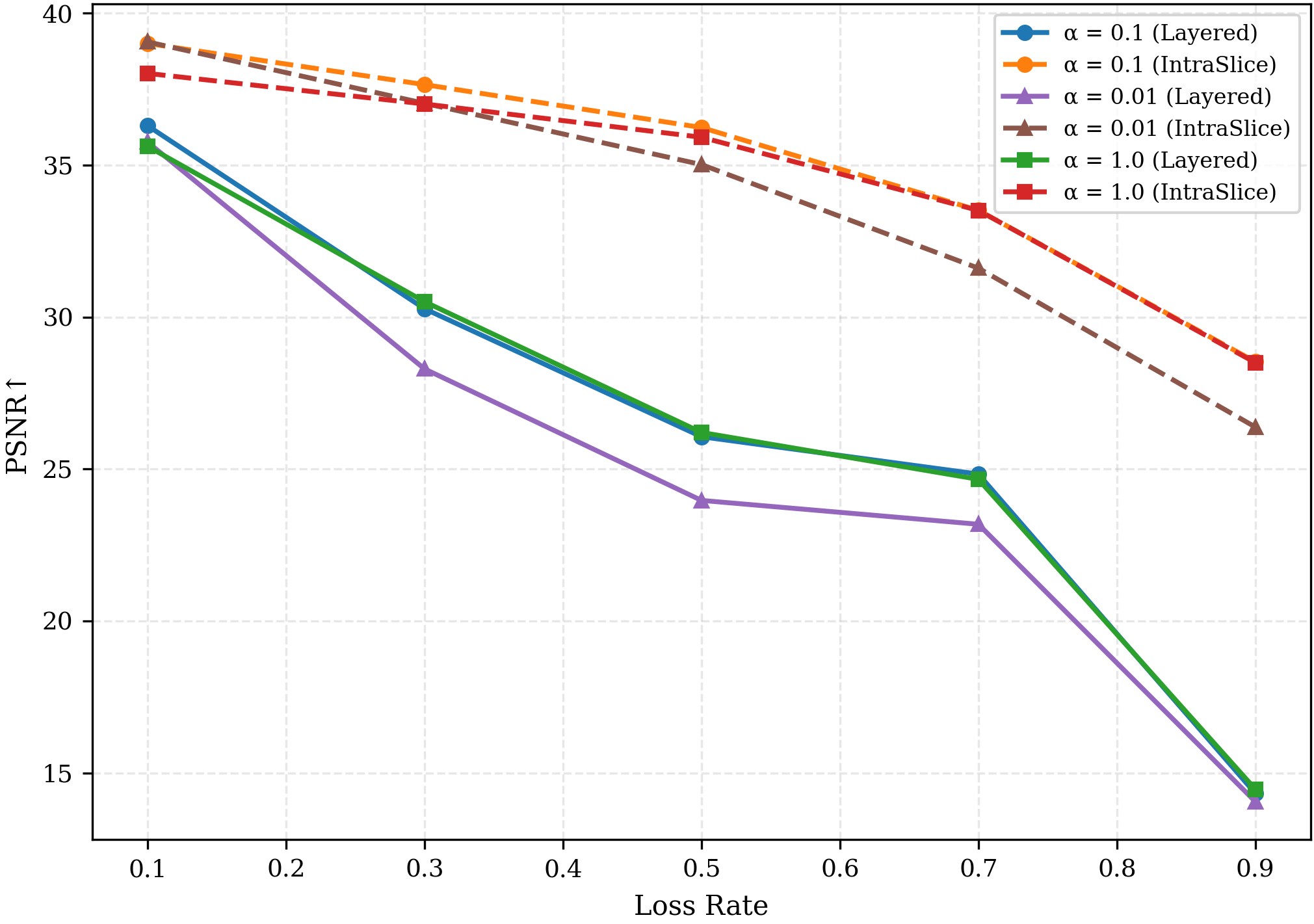}
    \caption{Effect of $\alpha$ in~\eqref{eq:d_total} on compression and loss resilience.}
    \label{fig:PSNR_vs_LossRate_alpha}
\end{figure}

Fig.~\ref{fig:PSNR_vs_pack_mode} shows the reconstruction performance with different packetization schemes. QLDS-based packetization, random packetization, and spatial-interleaving packetization achieve comparable performance under the intra-slice mode, where no interpacket reference dependency is introduced. This result indicates that effective packetization depends less on a particular sequence than on whether missing tokens are sufficiently dispersed over the two-dimensional latent map. By contrast, row-scan, row-interleaved, and Hilbert packetization retain pronounced spatial clustering or directional structure, increasing the likelihood of contiguous regions of missing tokens after a single packet loss and consequently degrading recovery performance. Although random packetization provides spatially dispersed packet allocation in expectation, its performance remains subject to randomness and may occasionally degrade sharply, as illustrated in Fig.~\ref{fig:PSNR_vs_pack_mode_b} and Fig.~\ref{fig:PSNR_vs_pack_mode_c}. As the segment length increases, interpacket dependencies cause a physical packet loss to propagate along the reference chain within a segment, resulting in an effective packet-loss rate substantially higher than the channel packet-loss rate and, consequently, a more rapid PSNR degradation. QLDS-based packetization achieves consistently strong performance across different coding modes. This advantage may stem from the nonperiodic, uniform coverage of the low-discrepancy sequence. Specifically, QLDS avoids the incidental local clustering of random packetization and mitigates the fixed periodic structure of regular-grid interleaving, making the spatial distribution of missing tokens more closely resemble the random-mask recovery target used during training.

Fig.~\ref{fig:PSNR_vs_LossRate_alpha} shows the effect of $\alpha$ in~\eqref{eq:d_total}, which balances compression performance against loss resilience. A smaller $\alpha$ places greater emphasis on reconstruction quality without packet loss than on transmission resilience. As the packet-loss rate increases from 0.1 to 0.9, the intra-slice mode with $\alpha=1$ exhibits a 3.16 dB smaller PSNR drop than that with $\alpha=0.01$. In the layered mode, the error-propagation chain is the longest, causing the reconstruction performance to degrade rapidly as the packet loss rate increases. When $\alpha=0.01$, the model exhibits limited robustness to packet loss, further resulting in inferior performance under this mode.

\begin{figure}[t]
    \centering
    \includegraphics[width=\linewidth]{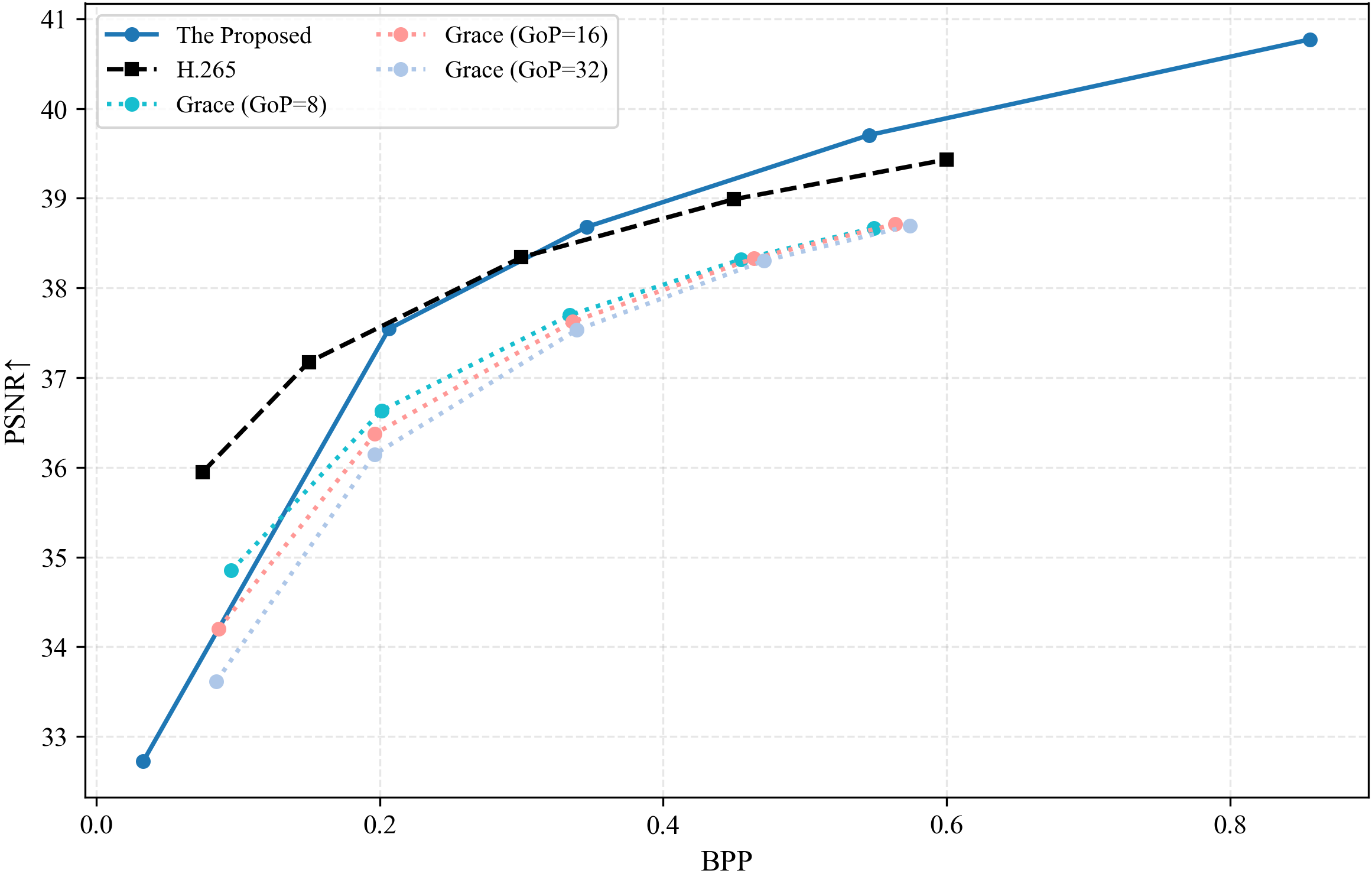}
    \caption{Reconstruction performance without packet loss.}
    \label{fig:NoLoss_PSNR_vs_BPP}
\end{figure}

\textbf{Loss-free performance and complexity:}
Fig.~\ref{fig:NoLoss_PSNR_vs_BPP} compares the reconstruction quality of ReLViC and the baselines in the absence of packet loss. It can be observed that GRACE improves transmission robustness at the cost of some compression efficiency. In contrast, ReLViC achieves compression performance comparable to, or even slightly better than, H.265 under loss-free conditions, while providing substantially greater resilience to packet loss than GRACE.

Table~\ref{tab:model_complexity} reports the model size and computational complexity. The model contains 231.155~M parameters, and its encoding and decoding costs are approximately 32.8~tera floating-point operations (TFLOPs). Spatiotemporal latent-context modeling accounts for most of this computation. On an RTX~4090, the model encodes a 1080p frame in approximately 0.8294~s, corresponding to approximately 1.2~frames per second (FPS).

\section{Conclusion}
\label{sec:conclusion}
This paper has presented ReLViC, a loss-resilient learned video coding framework that incorporates both latent recovery and packet-level organization. ReLViC couples compression and recovery through a dual-purpose Transformer, so that the spatiotemporal context learned for entropy modeling also provides the evidence needed to infer missing latent tokens. The progressive training procedure first establishes single-frame coding and then learns a temporal entropy model and the recovery of masked latent tokens. At the packet level, dispersed packetization prevents a single erasure from creating a large contiguous region of missing latent tokens, while controllable packet dependencies provide different trade-offs between compression efficiency and error-propagation range. In particular, selecting the segment length $g$ changes the maximum propagation span to $g-1$ packets without retraining. Experimental results demonstrate that ReLViC substantially outperforms H.265+FEC and GRACE under heavy packet loss and maintains favorable reconstruction quality at a packet-loss rate of $90\%$. Switching among packet-context modes enables ReLViC to balance compression efficiency and transmission resilience under different network conditions.

\bibliographystyle{IEEEtran}
\bibliography{ref}

@inproceedings{Rippel_2019_ICCV,
  author = {Rippel, Oren and Nair, Sanjay and Lew, Carissa and Branson, Steve and Anderson, Alexander G. and Bourdev, Lubomir},
  title = {Learned Video Compression},
  booktitle = {Proceedings of the IEEE/CVF International Conference on Computer Vision (ICCV)},
  pages = {3454--3463},
  month = oct,
  year = {2019},
  _url = {https://openaccess.thecvf.com/content_ICCV_2019/html/Rippel_Learned_Video_Compression_ICCV_2019_paper.html}
}

@inproceedings{Li_2023_CVPR,
  author = {Li, Jiahao and Li, Bin and Lu, Yan},
  title = {Neural Video Compression With Diverse Contexts},
  booktitle = {Proceedings of the IEEE/CVF Conference on Computer Vision and Pattern Recognition (CVPR)},
  pages = {22616--22626},
  month = jun,
  year = {2023},
  _url = {https://openaccess.thecvf.com/content/CVPR2023/html/Li_Neural_Video_Compression_With_Diverse_Contexts_CVPR_2023_paper.html}
}

@techreport{RFC4588,
  author = {Rey, J. and Leon, D. and Miyazaki, A. and Varsa, V. and Hakenberg, R.},
  title = {{RTP} Retransmission Payload Format},
  institution = {Internet Engineering Task Force},
  type = {Request for Comments},
  number = {4588},
  month = jul,
  year = {2006},
  doi = {10.17487/RFC4588},
  _url = {https://www.rfc-editor.org/rfc/rfc4588}
}

@article{Ahmed2021HARQ,
  author = {Ahmed, Ashfaq and Al-Dweik, Arafat and Iraqi, Youssef and Mukhtar, Husameldin and Naeem, Muhammad and Hossain, Ekram},
  title = {Hybrid Automatic Repeat Request ({HARQ}) in Wireless Communications Systems and Standards: A Contemporary Survey},
  journal = {IEEE Communications Surveys \& Tutorials},
  volume = {23},
  number = {4},
  pages = {2711--2752},
  year = {2021},
  doi = {10.1109/COMST.2021.3094401}
}

@inproceedings{Wang2026LADR,
  author = {Wang, Jing and Kong, Xiao and Ni, Yunzhe and Wen, Nian and Zhang, Jiaxing and Miao, Congcong and Liu, Honghao},
  title = {From Source to Solution: Tackling Packet Losses in Large-scale Cloud Gaming Systematically and Precisely},
  booktitle = {23rd USENIX Symposium on Networked Systems Design and Implementation (NSDI 26)},
  pages = {1191--1206},
  publisher = {USENIX Association},
  address = {Renton, WA},
  month = may,
  year = {2026},
  isbn = {978-1-939133-54-0},
  _url = {https://www.usenix.org/conference/nsdi26/presentation/wang-jing}
}

@techreport{RFC5109,
  author = {Li, A.},
  title = {{RTP} Payload Format for Generic Forward Error Correction},
  institution = {Internet Engineering Task Force},
  type = {Request for Comments},
  number = {5109},
  month = dec,
  year = {2007},
  doi = {10.17487/RFC5109},
  _url = {https://www.rfc-editor.org/rfc/rfc5109}
}

@inproceedings{Rudow2023Tambur,
  author = {Rudow, Michael and Yan, Francis Y. and Kumar, Abhishek and Ananthanarayanan, Ganesh and Ellis, Martin and Rashmi, K. V.},
  title = {Tambur: Efficient Loss Recovery for Videoconferencing via Streaming Codes},
  booktitle = {20th USENIX Symposium on Networked Systems Design and Implementation (NSDI 23)},
  pages = {953--971},
  publisher = {USENIX Association},
  address = {Boston, MA},
  month = apr,
  year = {2023},
  isbn = {978-1-939133-33-5},
  _url = {https://www.usenix.org/conference/nsdi23/presentation/rudow}
}

@article{Iliopoulos31122015,
author = {Vasileios Iliopoulos},
editor = {Yong Hong Wu},
title = {The plastic number and its generalized polynomial},
journal = {Cogent Mathematics},
volume = {2},
number = {1},
pages = {1023123},
year = {2015},
publisher = {Cogent OA},
doi = {10.1080/23311835.2015.1023123},
_URL = {https://doi.org/10.1080/23311835.2015.1023123},
eprint = { https://doi.org/10.1080/23311835.2015.1023123}
}

@article{Hadizadeh2011Burst,
  author = {Hadizadeh, Hadi and Bajic, Ivan V.},
  title = {Burst-Loss-Resilient Packetization of Video},
  journal = {IEEE Transactions on Image Processing},
  volume = {20},
  number = {11},
  pages = {3195--3206},
  month = nov,
  year = {2011},
  doi = {10.1109/TIP.2011.2132729}
}

@article{Chen2015Adaptive,
  author = {Chen, Haoming and Zhao, Chen and Sun, Ming-Ting and Drake, Aaron},
  title = {Adaptive Intra-Refresh for Low-Delay Error-Resilient Video Coding},
  journal = {Journal of Visual Communication and Image Representation},
  volume = {31},
  pages = {294--304},
  month = aug,
  year = {2015},
  doi = {10.1016/j.jvcir.2015.06.018}
}

@inproceedings{Benjak2021VVC,
  author = {Benjak, Martin and Samayoa, Yasser and Ostermann, J{\"o}rn},
  title = {Neural Network-Based Error Concealment for {VVC}},
  booktitle = {2021 IEEE International Conference on Image Processing (ICIP)},
  pages = {2114--2118},
  year = {2021},
  doi = {10.1109/ICIP42928.2021.9506399}
}

@inproceedings{Sivaraman2024Gemino,
  author = {Sivaraman, Vibhaalakshmi and Karimi, Pantea and Venkatapathy, Vedantha and Khani, Mehrdad and Fouladi, Sadjad and Alizadeh, Mohammad and Durand, Fr{\'e}do and Sze, Vivienne},
  title = {Gemino: Practical and Robust Neural Compression for Video Conferencing},
  booktitle = {21st USENIX Symposium on Networked Systems Design and Implementation (NSDI 24)},
  pages = {569--590},
  publisher = {USENIX Association},
  address = {Santa Clara, CA},
  month = apr,
  year = {2024},
  isbn = {978-1-939133-39-7},
  _url = {https://www.usenix.org/conference/nsdi24/presentation/sivaraman}
}

@article{RN2932,
  author = {Wang, Sixian and Dai, Jincheng and Qin, Xiaoqi and Yang, Ke and Niu, Kai and Zhang, Ping},
  title = {{ResiComp}: Loss-Resilient Image Compression via Dual-Functional Masked Visual Token Modeling},
  journal = {IEEE Transactions on Circuits and Systems for Video Technology},
  volume = {35},
  number = {7},
  pages = {7181-7195},
  ISSN = {1051-8215
1558-2205},
  doi = {10.1109/tcsvt.2025.3539747},
  year = {2025},
  type = {Journal Article}
}

@inproceedings{RN2730,
   author = {Cheng, Yihua and Zhang, Ziyi and Li, Hanchen and Arapin, Anton and Zhang, Yue and Zhang, Qizheng and Liu, Yuhan and Du, Kuntai and Zhang, Xu and Yan, Francis Y. and Mazumdar, Amrita and Feamster, Nick and Jiang, Junchen},
   title = {{GRACE}: Loss-Resilient Real-Time Video through Neural Codecs},
   booktitle = {21st USENIX Symposium on Networked Systems Design and Implementation (NSDI 24)},
   address = {Santa Clara, CA},
   publisher = {USENIX Association},
   pages = {509-531},
   year = {2024},
   _url = {https://www.usenix.org/conference/nsdi24/presentation/cheng},
   type = {Conference Proceedings},
   abstract={sion, two primary strategies are employed—encoder-based forward error correction (FEC) and decoder-based error concealment. The former encodes data with redundancy before transmission, yet determining the optimal redundancy level in advance proves challenging. The latter reconstructs video from partially received frames, but dividing a frame into independently coded partitions inherently compromises compression efficiency, and the lost information cannot be effectively recovered by the decoder without adapting the encoder. We present a loss-resilient real-time video system called
GRACE, which preserves the user’s quality of experience (QoE) across a wide range of packet losses through a new neural video codec. Central to GRACE’s enhanced loss resilience is its joint training ofthe neural encoder and decoder under a spectrum ofsimulated packet losses. In lossless scenarios, GRACE achieves video quality on par with conventional codecs (e.g., H.265). As the loss rate escalates, GRACE exhibits a more graceful, less pronounced decline in quality, consistently outperforming other loss-resilient schemes. Through extensive evaluation on various videos and real network traces, we demonstrate that GRACE reduces undecodable frames by 95\% and stall duration by 90\% compared with FEC, while markedly boosting video quality over error concealment methods. In a user study with 240 crowdsourced participants and 960 subjective ratings, GRACE registers a 38\% higher mean opinion score (MOS) than other baselines. We make the source codes and models of GRACE public at https://uchi-jcl.github.io/grace.html.}
}

@inproceedings{RN2680,
  author = {Li, Tianhong and Sivaraman, Vibhaalakshmi and Karimi, Pantea and Fan, Lijie and Alizadeh, Mohammad Sadegh and Katabi, Dina},
  title = {Reparo: Loss-Resilient Generative Codec for Video Conferencing},
  booktitle = {Proceedings of Machine Learning and Systems (MLSys)},
  year = {2026},
  pages = {1701--1718},
  volume = {8},
  _url = {https://vibhaalakshmi.com/publication/reparo/},
  abstract = {Packet loss during video conferencing often results in poor
quality and video freezing. Retransmitting lost packets is often impractical due to the need for real-time playback, and using Forward Error Correction (FEC) for packet recovery is challenging due to the unpredictable and bursty nature of Internet losses. Excessive redundancy leads to inefficiency and wasted bandwidth, while insufficient redundancy results in undecodable frames, causing video freezes and quality degradation in subsequent frames. We introduce Reparo — a loss-resilient video conferencing framework based on generative deep learning models to address these issues. Our approach generates missing information when a frame or part of a frame is lost. This generation is conditioned on the data received thus far, considering the model’s understanding of how people and objects appear and interact within the visual realm. Experimental results, using publicly available video conferencing datasets, demonstrate that Reparo outperforms state-of-the-art FEC-based video conferencing solutions in terms of both video quality (measured through PSNR, SSIM, and LPIPS) and the occurrence of video freezes.}
}

@article{RN2937,
  author = {Hu, Xinyue and Ye, Wei and Tang, Jiaxiang and Ramadan, Eman and Zhang, Zhi-Li},
  title = {Robust Multiple Description Neural Video Codec with Masked Transformer for Dynamic and Noisy Networks},
  journal = {arXiv preprint arXiv:2412.07922},
  year = {2024},
  type = {Journal Article}
}

@inproceedings{10.1145/3339825.3394937,
author = {Mercat, Alexandre and Viitanen, Marko and Vanne, Jarno},
title = {{UVG} dataset: 50/120fps {4K} sequences for video codec analysis and development},
year = {2020},
isbn = {9781450368452},
publisher = {Association for Computing Machinery},
address = {New York, NY, USA},
_url = {https://doi.org/10.1145/3339825.3394937},
doi = {10.1145/3339825.3394937},
abstract = {This paper provides an overview of our open Ultra Video Group (UVG) dataset that is composed of 16 versatile 4K (3840\texttimes{}2160) test video sequences. These natural sequences were captured either at 50 or 120 frames per second (fps) and stored online in raw 8-bit and 10-bit 4:2:0 YUV formats. The dataset is published on our website (ultravideo.cs.tut.fi) under a non-commercial Creative Commons BY-NC license. In this paper, all UVG sequences are described in detail and characterized by their spatial and temporal perceptual information, rate-distortion behavior, and coding complexity with the latest HEVC/H.265 and VVC/H.266 reference video codecs. The proposed dataset is the first to provide complementary 4K sequences up to 120 fps and is therefore particularly valuable for cutting-edge multimedia applications. Our evaluations also show that it comprehensively complements the existing 4K test set in VVC standardization, so we recommend including it in subjective and objective quality assessments of next-generation VVC codecs.},
booktitle = {Proceedings of the 11th ACM Multimedia Systems Conference},
pages = {297–-302},
numpages = {6},
keywords = {versatile video coding (VVC), ultra high definition (UHD), raw video, open dataset, high efficiency video coding (HEVC)},
location = {Istanbul, Turkey},
series = {MMSys '20}
}

@inproceedings{milner2004analysis,
  title={An Analysis of Packet Loss Models for Distributed Speech Recognition},
  author={Milner, BP and James, AB},
  booktitle={8th International Conference on Spoken Language Processing (Interspeech 2004)},
  pages={1549--1552},
  year={2004}
}

@INPROCEEDINGS{9879846,
  author={He, Dailan and Yang, Ziming and Peng, Weikun and Ma, Rui and Qin, Hongwei and Wang, Yan},
  booktitle={2022 IEEE/CVF Conference on Computer Vision and Pattern Recognition (CVPR)}, 
  title = {{ELIC}: Efficient Learned Image Compression with Unevenly Grouped Space-Channel Contextual Adaptive Coding},
  year={2022},
  volume={},
  number={},
  pages={5708-5717},
  doi ={10.1109/CVPR52688.2022.00563}}

@INPROCEEDINGS{749301,
  author={Yajnik, M. and Sue Moon and Kurose, J. and Towsley, D.},
  booktitle={IEEE INFOCOM '99. Conference on Computer Communications. Proceedings. Eighteenth Annual Joint Conference of the IEEE Computer and Communications Societies. The Future is Now (Cat. No.99CH36320)}, 
  title = {Measurement and modelling of the temporal dependence in packet loss},
  year={1999},
  volume={1},
  number={},
  pages={345-352 vol.1},
  doi ={10.1109/INFCOM.1999.749301}}

@ARTICLE{1510488,
  author={Carvalho, L. and Angeja, J. and Navarro, A.},
  journal={IEEE Transactions on Consumer Electronics}, 
  title = {A new packet loss model of the {IEEE} 802.11g wireless network for multimedia communications},
  year={2005},
  volume={51},
  number={3},
  pages={809-814},
  doi ={10.1109/TCE.2005.1510488}}

@ARTICLE{664283,
  author={Yao Wang and Qin-Fan Zhu},
  journal={Proceedings of the IEEE}, 
  title = {Error control and concealment for video communication: a review},
  year={1998},
  volume={86},
  number={5},
  pages={974-997},
  doi ={10.1109/5.664283}}

@ARTICLE{855913,
  author={Yao Wang and Wenger, S. and Jiantao Wen and Katsaggelos, A.K.},
  journal={IEEE Signal Processing Magazine}, 
  title = {Error resilient video coding techniques},
  year={2000},
  volume={17},
  number={4},
  pages={61-82},
  doi ={10.1109/79.855913}}

@inproceedings{meng2024hairpin,
  author = {Meng, Zili and Kong, Xiao and Chen, Jing and Wang, Bo and Xu, Mingwei and Han, Rui and Liu, Honghao and Arun, Venkat and Hu, Hongxin and Wei, Xue},
  title = {Hairpin: Rethinking Packet Loss Recovery in Edge-Based Interactive Video Streaming},
  booktitle = {21st USENIX Symposium on Networked Systems Design and Implementation (NSDI 24)},
  address = {Santa Clara, CA},
  publisher = {USENIX Association},
  pages = {907--926},
  year = {2024},
  month = apr,
  _url = {https://www.usenix.org/conference/nsdi24/presentation/meng},
  abstract={Interactive streaming requires minimizing stuttering events (or deadline misses for video frames) to ensure seamless interaction between users and applications. However, existing packet loss recovery mechanisms uniformly optimize redundancy for initial transmission and retransmission, which still could not satisfy the delay requirements of interactive streaming, but also introduces considerable bandwidth costs. Our insight is that in edge-based interactive streaming, differentiating retransmissions on redundancy settings can often achieve a low bandwidth cost and a low deadline miss rate simultaneously. In this paper, we propose Hairpin, a new packet loss recovery mechanism for edge-based interactive streaming. Hairpin finds the optimal combination of data packets, retransmissions, and redundant packets over multiple rounds of transmissions, which significantly reduces the bandwidth cost while ensuring the end-to-end latency requirement. Experiments with production deployments demonstrate that Hairpin can simultaneously reduce the bandwidth cost by 40% and the deadline miss rate by 32% on average in the wild against state-of-the-art solutions.}
}

@article{RN3030,
   author = {Liu, Zhenyu and Ma, Yi and Tafazolli, Rahim and Ding, Zhi},
   title = {{Resi-VidTok}: An Efficient and Decomposed Progressive Tokenization Framework for Ultra-Low-Rate and Lightweight Video Transmission},
   journal = {arXiv preprint arXiv:2510.25002},
   year = {2025},
   type = {Journal Article},
}

@article{RN2456,
  author = {Tung, Tze-Yang and Gunduz, Deniz},
  title = {{DeepWiVe}: Deep-Learning-Aided Wireless Video Transmission},
  journal = {IEEE Journal on Selected Areas in Communications},
  volume = {40},
  number = {9},
  pages = {2570-2583},
  ISSN = {0733-8716
1558-0008},
  doi = {10.1109/jsac.2022.3191354},
  year = {2022},
  type = {Journal Article}
}

@article{RN2798,
   author = {Wang, Tao and Zheng, Zhigao and Lin, Yun and Yao, Shihong and Xie, Xiao},
   title = {Reliable and Robust Unmanned Aerial Vehicle Wireless Video Transmission},
   journal = {IEEE Transactions on Reliability},
   volume = {68},
   number = {3},
   pages = {1050-1060},
   ISSN = {0018-9529
1558-1721},
   DOI = {10.1109/tr.2018.2864683},
   year = {2019},
   type = {Journal Article},
}

@article{RN2195,
   author = {Wu, Jiyan and Yuen, Chau and Cheung, Ngai-Man and Chen, Junliang and Chen, Chang Wen},
   title = {Streaming Mobile Cloud Gaming Video Over {TCP} With Adaptive Source–{FEC} Coding},
   journal = {IEEE Transactions on Circuits and Systems for Video Technology},
   volume = {27},
   number = {1},
   pages = {32-48},
   ISSN = {1051-8215
1558-2205},
   DOI = {10.1109/tcsvt.2016.2527398},
   year = {2017},
   type = {Journal Article},
}

@inproceedings {316606,
author = {Tianyi Gong and Zijian Cao and Zixing Zhang and Jiangkai Wu and Xinggong Zhang and Shuguang Cui and Fangxin Wang},
title = {Morphe: {High-Fidelity} Generative Video Streaming with Vision Foundation Model},
booktitle = {23rd USENIX Symposium on Networked Systems Design and Implementation (NSDI 26)},
year = {2026},
isbn = {978-1-939133-54-0},
address = {Renton, WA},
pages = {301--317},
_url = {https://www.usenix.org/conference/nsdi26/presentation/gong},
publisher = {USENIX Association},
abstract={Video streaming is a fundamental Internet service, while
the quality still cannot be guaranteed especially in poor network conditions such as bandwidth-constrained and remote areas. Existing works mainly work towards two directions: traditional pixel-codec streaming nearly approaches its limit and is hard to step further in compression; the emerging neuralenhanced or generative streaming usually fall short in latency and visual fidelity, hindering their practical deployment. Inspired by the recent success of vision foundation model
(VFM), we strive to harness the powerful video understanding and processing capacities ofVFM to achieve generalization, high fidelity and loss resilience for real-time video streaming with even higher compression rate. We present Morphe1, the first revolutionized paradigm that enables VFM-based endto-end generative video streaming towards this goal. Specifically, Morphe employs joint training of visual tokenizers and variable-resolution spatiotemporal optimization under simulated network constraints. Additionally, a robust streaming system is constructed that leverages intelligent packet dropping to resist real-world network perturbations. Extensive evaluation demonstrates that Morphe achieves comparable visual quality while saving 62.5% bandwidth compared to H.265, and accomplishes real-time, loss-resilient video delivery in challenging network environments, representing a milestone in VFM-enabled multimedia streaming solutions.},
}

@article{Sheng2023TCM,
  author = {Sheng, Xihua and Li, Jiahao and Li, Bin and Li, Li and Liu, Dong and Lu, Yan},
  title = {Temporal Context Mining for Learned Video Compression},
  journal = {IEEE Transactions on Multimedia},
  volume = {25},
  pages = {7311--7322},
  year = {2023},
  doi = {10.1109/TMM.2022.3220421}
}

@inproceedings{Li2022HEM,
  author = {Li, Jiahao and Li, Bin and Lu, Yan},
  title = {Hybrid Spatial-Temporal Entropy Modelling for Neural Video Compression},
  booktitle = {Proceedings of the 30th ACM International Conference on Multimedia},
  pages = {1503--1511},
  year = {2022},
  doi = {10.1145/3503161.3547845}
}

@inproceedings{Wang2023EVC,
  author = {Wang, Guo-Hua and Li, Jiahao and Li, Bin and Lu, Yan},
  title = {{EVC}: Towards Real-Time Neural Image Compression with Mask Decay},
  booktitle = {International Conference on Learning Representations},
  year = {2023},
  _url = {https://openreview.net/forum?id=XUxad2Gj40n}
}

@inproceedings{RN2940,
   author = {Li, Jiahao and Li, Bin and Lu, Yan},
   title = {Deep contextual video compression},
   booktitle = {Advances in Neural Information Processing Systems},
   volume = {34},
   pages = {18114-18125},
   year = {2021},
   type = {Conference Proceedings}
}

@INPROCEEDINGS{10655044,
  author={Li, Jiahao and Li, Bin and Lu, Yan},
  booktitle={2024 IEEE/CVF Conference on Computer Vision and Pattern Recognition (CVPR)}, 
  title={Neural Video Compression with Feature Modulation}, 
  year={2024},
  volume={},
  number={},
  pages={26099-26108},
  doi={10.1109/CVPR52733.2024.02466}}

@inproceedings{Jia_2025_CVPR,
    author    = {Jia, Zhaoyang and Li, Bin and Li, Jiahao and Xie, Wenxuan and Qi, Linfeng and Li, Houqiang and Lu, Yan},
    title     = {Towards Practical Real-Time Neural Video Compression},
    booktitle = {Proceedings of the IEEE/CVF Conference on Computer Vision and Pattern Recognition (CVPR)},
    month     = {June},
    year      = {2025},
    pages     = {12543-12552},
    doi       = {10.1109/CVPR52734.2025.01170},
}

@article{wiegand2003avc,
  author = {Wiegand, Thomas and Sullivan, Gary J. and Bjontegaard, Gisle and Luthra, Ajay},
  title = {Overview of the {H}.264/{AVC} Video Coding Standard},
  journal = {IEEE Transactions on Circuits and Systems for Video Technology},
  volume = {13},
  number = {7},
  pages = {560--576},
  year = {2003},
  doi = {10.1109/TCSVT.2003.815165}
}

@article{sullivan2012hevc,
  author = {Sullivan, Gary J. and Ohm, Jens-Rainer and Han, Woo-Jin and Wiegand, Thomas},
  title = {Overview of the High Efficiency Video Coding ({HEVC}) Standard},
  journal = {IEEE Transactions on Circuits and Systems for Video Technology},
  volume = {22},
  number = {12},
  pages = {1649--1668},
  year = {2012},
  doi = {10.1109/TCSVT.2012.2221191}
}

@article{lin1984arq,
  author = {Lin, Shu and Costello, Daniel J. and Miller, Michael J.},
  title = {Automatic-Repeat-Request Error-Control Schemes},
  journal = {IEEE Communications Magazine},
  volume = {22},
  number = {12},
  pages = {5--17},
  year = {1984},
  doi = {10.1109/MCOM.1984.1091865}
}

@inproceedings{lu2019dvc,
  author = {Lu, Guo and Ouyang, Wanli and Xu, Dong and Zhang, Xiaoyun and Cai, Chunlei and Gao, Zhiyong},
  title = {{DVC}: An End-to-End Deep Video Compression Framework},
  booktitle = {Proceedings of the IEEE/CVF Conference on Computer Vision and Pattern Recognition},
  year = {2019}
}

@article{bross2021vvc,
  author = {Bross, Benjamin and Chen, Jianle and Liu, Shan and Wang, Ye-Kui},
  title = {Overview of the Versatile Video Coding ({VVC}) Standard and its Applications},
  journal = {IEEE Transactions on Circuits and Systems for Video Technology},
  volume = {31},
  number = {10},
  pages = {3736--3764},
  year = {2021},
  doi = {10.1109/TCSVT.2021.3101953}
}

@article{AFZAL2023103581,
title = {A holistic survey of multipath wireless video streaming},
journal = {Journal of Network and Computer Applications},
volume = {212},
pages = {103581},
year = {2023},
issn = {1084-8045},
doi = {https://doi.org/10.1016/j.jnca.2022.103581},
_url = {https://www.sciencedirect.com/science/article/pii/S1084804522002223},
author = {Samira Afzal and Vanessa Testoni and Christian Esteve Rothenberg and Prakash Kolan and Imed Bouazizi},
keywords = {Wireless video streaming, Multipath routing, Packet scheduling, Heterogeneous networks},
abstract = {Demand for wireless video streaming services increases with users expecting to access high-quality video streaming experiences. Ensuring Quality of Experience (QoE) is quite challenging due to varying bandwidth and time constraints. Since most of today’s mobile devices are equipped with multiple network interfaces, one promising approach is to benefit from multipath communications. Multipathing leads to higher aggregate bandwidth and distributing video traffic over multiple network paths improves stability, seamless connectivity, and QoE. However, most of current transport protocols do not match the requirements of video streaming applications or are not designed to address relevant issues, such as networks heterogeneity, head-of-line blocking, and delay constraints. In this comprehensive survey, we first review video streaming standards and technology developments. We then discuss the benefits and challenges of multipath video transmission over wireless. We provide a holistic literature review of multipath wireless video streaming, shedding light on the different alternatives from an end-to-end layered stack perspective, reviewing key multipath wireless scheduling functions, unveiling trade-offs of each approach, and presenting a suitable taxonomy to classify the state-of-the-art. Finally, we discuss open issues and avenues for future work.}
}

@article{10.1145/3607139,
author = {Fujihashi, Takuya and Koike-Akino, Toshiaki and Watanabe, Takashi},
title = {Soft Delivery: Survey on a New Paradigm for Wireless and Mobile Multimedia Streaming},
year = {2023},
issue_date = {February 2024},
publisher = {Association for Computing Machinery},
address = {New York, NY, USA},
volume = {56},
number = {2},
issn = {0360-0300},
_url = {https://doi.org/10.1145/3607139},
doi = {10.1145/3607139},
abstract = {The increasing demand for video streaming services is the key driver of modern wireless and mobile communications. Although many studies have designed digital-based delivery schemes to send video content over wireless and mobile networks, significant quality degradation, known as cliff and leveling effects, often occurs owing to fluctuating channel characteristics. In this article, we present a comprehensive summary of soft delivery, which is a new paradigm for wireless and mobile video streaming and discuss the future directions of soft delivery. Existing studies found that introducing multi-dimensional cosine transform, human vision system, and graph signal processing can make soft delivery schemes more effective in untethered immersive experiences, including virtual reality and volumetric media, than digital-based delivery schemes. In addition, this study finds that soft delivery has the potential to be a new standard to deliver deep neural network models and tactile information over wireless and mobile networks.},
journal = {ACM Comput. Surv.},
month = sep,
articleno = {33},
numpages = {37},
keywords = {Soft delivery, hybrid digital–analog delivery, extended reality}
}

@article{BAENA2023109808,
title = {Measuring and estimating Key Quality Indicators in Cloud Gaming services},
journal = {Computer Networks},
volume = {231},
pages = {109808},
year = {2023},
issn = {1389-1286},
doi = {https://doi.org/10.1016/j.comnet.2023.109808},
_url = {https://www.sciencedirect.com/science/article/pii/S1389128623002530},
author = {Carlos Baena and O.S. Peñaherrera-Pulla and Raquel Barco and Sergio Fortes},
keywords = {Cloud gaming, Mobile networks, Quality of experience, Key Quality Indicators, End-to-end, Machine learning},
abstract = {The gaming industry has proposed the concept of Cloud Gaming (CG), a paradigm that enhances the gaming experience on reduced hardware devices. However, this paradigm puts a lot of pressure on the communication links that connect the user to the cloud. As a result, the service experience becomes highly dependent on network connectivity. In this context, the present work proposes a framework for measuring and estimating the most important E2E (end-to-end) metrics of the CG service, namely Key Quality Indicators (KQIs). Therefore, different machine learning (ML) techniques are evaluated to predict KQIs related to the CG user experience. For this purpose, the most important KQIs of the service, such as input lag, freezes or perceived video frame rate, are collected in a real network deployment. The results show that ML techniques can be used to estimate these indicators solely from network-related metrics. This is seen as a valuable asset for the delivery of CG services over cellular networks, even without access to the user’s device, as it is expected for telecom operators.}
}

@ARTICLE{10668820,
  author={Sharma, Mohit K. and Farhat, Ibrahim and Liu, Chen-Feng and Sehad, Nassim and Hamidouche, Wassim and Debbah, Mérouane},
  journal={IEEE Open Journal of the Communications Society}, 
  title={Real-Time Immersive Aerial Video Streaming: A Comprehensive Survey, Benchmarking, and Open Challenges}, 
  year={2024},
  volume={5},
  number={},
  pages={5680-5705},
  doi={10.1109/OJCOMS.2024.3455763}}

\end{document}